\documentclass[preprint,preprintnumbers,amsmath,amssymb]{revtex4}

\usepackage{graphicx}
\usepackage{dcolumn}
\usepackage{bm}
\usepackage[german,american]{babel}

\newcommand{\figref}[1]{Fig.\ \ref{#1}}
\newcommand{\hyp}[2]{\mbox{#1 $-$ #2}}

\begin{document}

\title{Understanding correlation effects for ion conduction in polymer electrolytes}

\author{Arijit Maitra}
\author{Andreas Heuer}%
\email{arijitmaitra@uni-muenster.de,andheuer@uni-muenster.de}

\affiliation{
Westf\"{a}lische Wilhelms-Universit\"{a}t M\"{u}nster, Institut f\"{u}r Physikalische Chemie, Corrensstr. 30, 48149 M\"{u}nster, Germany }
\affiliation{ NRW Graduate School of Chemistry, Corrensstr. 36, 48149 M\"{u}nster, Germany}

\date{\today}

\begin{abstract}

Polymer electrolytes typically exhibit diminished ionic conductivity due to the presence of correlation effects between the cations and anions. Microscopically, transient ionic aggregates, e.g.\ {\it ion-pairs}, {\it ion-triplets} or higher order ionic clusters, engender ionic correlations. Employing {\it all-atom} simulation of a model polymer electrolyte comprising of poly(ethylene oxide) and lithium iodide, the ionic correlations are explored through construction of elementary functions between pairs of the ionic species that qualitatively explains the spatio-temporal nature of these correlations. Furthermore, commencing from the exact Einstein-like equation describing the collective diffusivity of the ions in terms of the average diffusivity of the ions (i.e. the self terms) and the correlations from distinct pairs of ions, several phenomenological parameters are introduced to keep track of the simplification procedure that finally boils down to the recently proposed phenomenological model by Stolwijk-Obeidi (SO) [N. A. Stolwijk and S. Obeidi, Phys. Rev. Lett. 93, 125901, 2004]. The approximation parameters, which can be retrieved from simulations, point to the necessity of additional information in order to fully describe the correlation effects apart from merely the fraction of ion-pairs which apparently accounts for the correlations originating from only the nearest neighbor structural correlations. These parameters are close to but not exactly unity as assumed in the SO model. Finally, as an application of the extended SO model one is able to estimate the dynamics of the free and non-free ions as well as their fractions from the knowledge of the single particle diffusivities and the collective diffusivity of the ions.
\end{abstract}

\maketitle

\section{\bf Introduction}

Polymer  electrolytes \cite{RatnerRev,BruceVin93, ScrosatiMRS00, RatnerMRS}
have received increasing attention for electrolytic applications in
energy storing devices like batteries. Cations, ideally lithium
ions (Li$^+$),  supplied by a dissociated salt in a polymer medium
provide for the charge conduction. It is known from experiments
and computer simulations that the collective diffusivity (which is
directly related to the ionic conductivity) in polymer
electrolytes falls short of the average diffusivity of the charge
carrying species (i.e. cations and anions). The reason is
attributed to the presence of motional correlations between the
unlike charges.

Molecular dynamics simulation \cite{Allen,FrenkelSmit} of an amorphous polymer electrolyte
offers the possibility of exploring the nature of ionic
correlations and their repercussions on the ionic conductivity.
During the past decade realistic simulations on linear chains of
poly(ethylene oxide) (PEO) with lithium salt have yielded insight
into the ionic mobilities and conductivities \cite{Plathe,Plathe2,
Neyertz95,BoroLiIStatic,BoroLiIDynamic,PEOLiPF6,BoroMDLiBF4,BoroMDLiTFSI}. Typically, cations
in polymer electrolytes can exist with or without the presence of
anions in their vicinity. Cations which are coordinated solely by
the ether oxygen (EO) atoms of the polymer host (i.e. PEO) are the
so called {\it free} carriers and their proportion is important
for the magnitude of ionic conductivity. However, a cation can
also attract one, two or more anionic neighbors around itself
forming the so called {\it pairs}, {\it triplets} or higher order
{\it clusters}. The distribution of the oxygen and the anionic
coordination numbers is a consequence of the rich interplay
between the interionic interactions and polymer entropy. Presence
of ionic aggregates, specifically the neutral ones like the {\it
pairs}, will not contribute to conductivity.

The average diffusivity of the ions, $D_{avg}$ is related
to the ionic  conductivity, $\sigma$, in an electrolyte at
temperature T by the equation
\begin{subequations}\label{nernst}
\begin{align}
 D_{avg} &= \sum_s D_s x_s \equiv H D_\sigma  \label{nernst1} \\
\mbox{where } & D_\sigma =  \sigma \frac{k_B T V}{e^2 n}. \label{nernst2}
\end{align}
\end{subequations}
Here $D_s$ and $x_s$ are the
diffusivity and the ionic fraction of species $s$, respectively;
$H$ the Haven ratio,  $D_\sigma$ \cite{note4} the collective diffusivity;
and $n$ is the total number of ions (both cations and anions),
in a volume $V$. With $H=1$ Eq.\ \eqref{nernst1} expresses the Nernst-Einstein equation.
$D_\sigma$ can be found from conductivity experiments \cite{Stolwijk04,ObeidiMacro,PasBan06,Akgol07}
whereas $D_s$ can be obtained from tracer diffusion measurements
\cite{Stolwijk04,ObeidiMacro} or PFG-NMR \cite{Hayamizu1,Hayamizu2}. Without ionic correlations the
Haven ratio $H$ is unity, yielding
 $D_\sigma = D_{avg}$. If ionic correlations are present this relation
needs to be augmented by the contributions from the cross
correlation terms between the distinct ions, yielding $H \ne 1$.
Correlated ionic motion can cause either $H<1$ (as in inorganic
ionic conductors) or $H>1$ (as in polymer electrolytes). Note that
in the latter case several cross correlations are present
(cation-cation, anion-anion and cation-anion).

Recently Stolwijk and Obeidi proposed a model (SO-model) that aims
to explain the depression of ionic conductivity relative to the
average ionic diffusivity in a representative electrolyte system
of PEO and sodium iodide (NaI) \cite{Stolwijk04}. This model is an attempt
to determine the mobilities of free ions and ion-pairs from the
experimental information comprising of the tracer diffusivities of
the ions and the ionic conductivity of the polymer electrolyte
system. The key assumption of the model is that the ions exist
either as free Na$^+$ and free I$^-$ or in the form of contact
ion-pairs (Na$^+$ $-$ I$^-$). 
Though the ion-pairs contribute to
mass transport, they do not participate in charge transport due to
charge neutrality, thereby, resulting in a diminished ionic
conductivity. Their model predicts that, over a broad temperature range,
the ion-pairs are orders of magnitude faster than either of
the free ionic species \cite{Stolwijk04}. This led them to interpret
that the ion-pairs are not coordinated with the ether oxygen atoms.
In light of the
insights from computer simulations where most of the cations (both
free and those associated with anions) are structurally
correlated and dynamically coupled to the ether oxygen atoms
\cite{MHprl} the SO model can be extended to accommodate the scenario
where the free ions and ion-pairs should display
similar mobility.

In this work we explore in detail the microscopic characteristics
of the cross correlation contribution in a polymer melt via
computer simulations for PEO with lithium iodide (LiI) as a salt. Our goal is
fourfold. First, we derive an exact formal equation for
$D_{\sigma}$ [viz. Eq.\ \eqref{kubo2}] which fully contains
the effect of dynamic correlations. Second, these correlations are
interpreted in terms of simple physical pictures. For example it
will be shown that ionic correlations exist beyond the nearest
neighbor shell which implies that structural ion-pairs, i.e. those
which are nearest neighbors,  are not sufficient to grasp
correlation effects. Third, we introduce a number of
approximations such that the exact equation boils down to the
phenomenological SO-model together with a few phenomenological
parameters (all close to but not exactly one) by which we keep
track of the approximations. For the present system, the values of
these parameters can be retrieved from the simulations. This offers
the possibility of turning the SO-model more general and accurate.
This attempt also brings out the necessity of additional
information apart from merely the tracer diffusivities and ionic
conductivity to quantify the microscopic dynamics. Fourth, we
apply the extended SO-model to the experimental data, reported in
\cite{Stolwijk04}. It turns out that the data can be explained without
invoking significant dynamical differences between the free ions and
ion-pairs. The resulting fraction of associated ions from the model
can be compared against the experimentally obtained information about the proportions of structural pairs via Raman scattering techniques. \\

\section{\bf Simulation Method}

\begin{figure}
  \includegraphics[clip,width=0.6\linewidth]{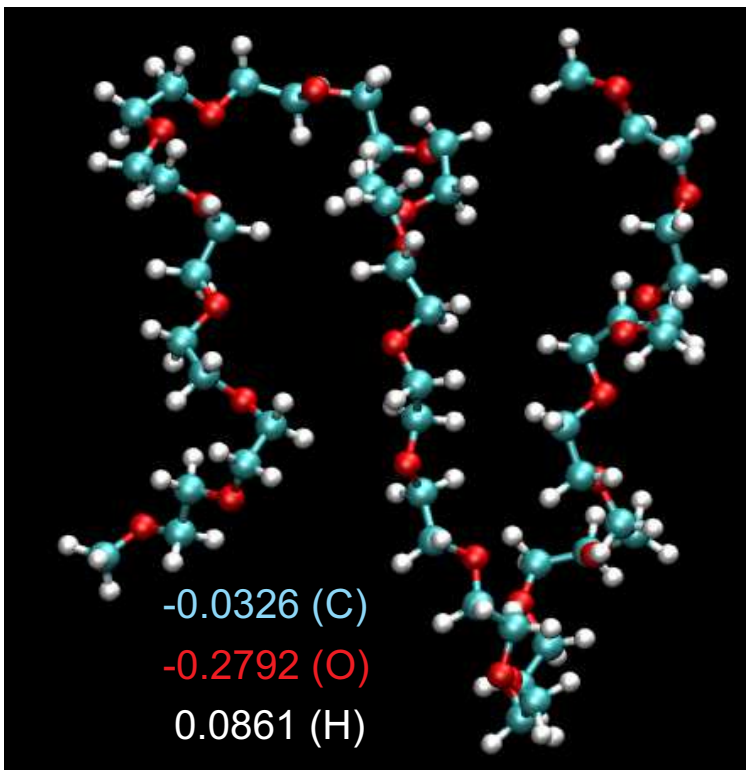}
  \caption{\label{PEO} The atomistic model of poly(ethylene oxide) with the partial charges 
   used are -0.0326 for carbon (cyan), -0.2792 for oxygen (red), 0.0861 for
   hydrogen (white) and -0.1187 for the terminal carbon atoms \cite{PEOFF2003}.}
\end{figure}

We have performed {\it all-atom} molecular dynamics  simulation
\cite{Allen,FrenkelSmit} on a system of a polymer electrolyte
comprising of chains of PEO and LiI as the salt
in the canonical ensemble (NVT) using the GROMACS \cite{gro}
package. The salt concentration in terms of the number of EO units
and Li$^+$ ions in the system was EO:Li$^+$=20:1. Conventional
periodic boundary conditions were applied to get rid of surface
effects. The simulation box contained 32 PEO chains. Each of the
polymer chains had the chemical formula H(CH$_2$$-$O$-$CH$_2$)$_N$H
where $N$ is the number of repeat units or chain length (Fig.\ \ref{PEO}). In this
work we have chosen $N=24$ which is a reasonable chain length to
study the dynamics of the ions/polymers within a production run
time window of $\approx$ 85 ns following an equlibration time of
about 10 ns. The Nose-Hoover thermostat \cite{Nose,hoover} algorithm was
set to maintain an average temperature throughout the simulation
run. The temperatures chosen were T = 425 K and T = 450 K. The
relaxation time of the PEO chains in terms of the Rouse time,
calculated by fitting the Rouse model \cite{DoiEd} to the dynamics of the
oxygen atoms were about 9.5 and 4.3 ns at T = 425 K and T = 450 K,
respectively. These relatively high temperatures were necessary to
achieve equilibration and obtain reasonable estimates, in
particular, for the collective diffusivity.  Prior to the NVT runs the system density had
been adjusted via an NPT (isobaric-isothermal ensemble with constant
number of particles) run for durations of about 5 ns with an average pressure of 1
MPa. 

The starting configuration 
of the polymer melt for the NPT run was
picked up from a thoroughly equilibrated system of neat PEO at a
temperature of 500 K. The ions were then randomly dispersed in it (i.e.\ 
positions of the ions were randomly selected) 
such that the minimum distance between a newly placed ion and  
the already existing particles (i.e.\ previously placed ions as well as atoms of the polymer) in
the system was about 3 \AA. To obtain a stable configuration, the system was then 
energy-minimised using the steepest descent technique \cite{gro}.
The subsequent NPT runs 
produced converged average densities of 1082 gcm$^{-3}$ and 1099 gcm$^{-3}$
at T = 450 K and T = 425 K respectively. The density fluctuations were of the 
order of 3.5 gcm$^{-3}$. 
Following the NPT runs, the systems were further equilibrated at constant
density and temperature (NVT) for more than 5 ns.    
In this paper we restrict ourselves
to the \mbox{T = 450 K} data because of the better statistics.

The force field for the description of intra- (bond lengths,
bend-angle and  dihedral potentials parameters) and
inter-molecular (e.g.\ Buckingham potential parameters, partial
charges) interactions of the PEO and the interactions between the
PEO and the Li$^+$ ions are taken from Refs.\
\cite{PEOFF2003,BoroMDLiBF4}. The interactions between the I$^-$
ions and the PEO and between the ions are essentially from Ref.\
\cite{PEOLiIFF97} with slight modifications \cite{interact}.
The attractive part of the Buckingham potential
between the \hyp{I}{Li}, \hyp{I}{I} and \hyp{I}{PEO} have been
slightly diminished. However, the charges ($+$1$e$ and $-$1$e$ for
Li$^+$ and I$^-$, respectively) and the Coulomb interactions between
the ions were unaltered and
calculated using the particle-mesh Ewald technique within a
distance cut-off of 10 \AA.
This was done in order to
generate a higher number of isolated {\it ion-pairs} and {\it
ion-triplets}. However, the fraction of free and non-free ions in the system
remained the same.
Notwithstanding, with this modification the equilibrium polymer
conformation of the sequence of dihedrals for the triad
O$-$C$-$C$-$O was found to be in agreement with the published
results of Ref.\ \cite{BoroLiIStatic} for their system of PEO/LiI
with a composition of EO:Li = 15:1 at T = 450 K. The qualitative
nature of the ionic correlations reported later in this article
also turned out to be similar for both the original and the modified potential. In short,
introducing the changes made to the potential only affected the
state of ionic association without involving the average polymer
conformations and dynamics. \\

\section{\bf Self Diffusion and Collective Diffusion}

\begin{figure}
  \vspace{5pt}
  \includegraphics[clip,width=0.9\linewidth]{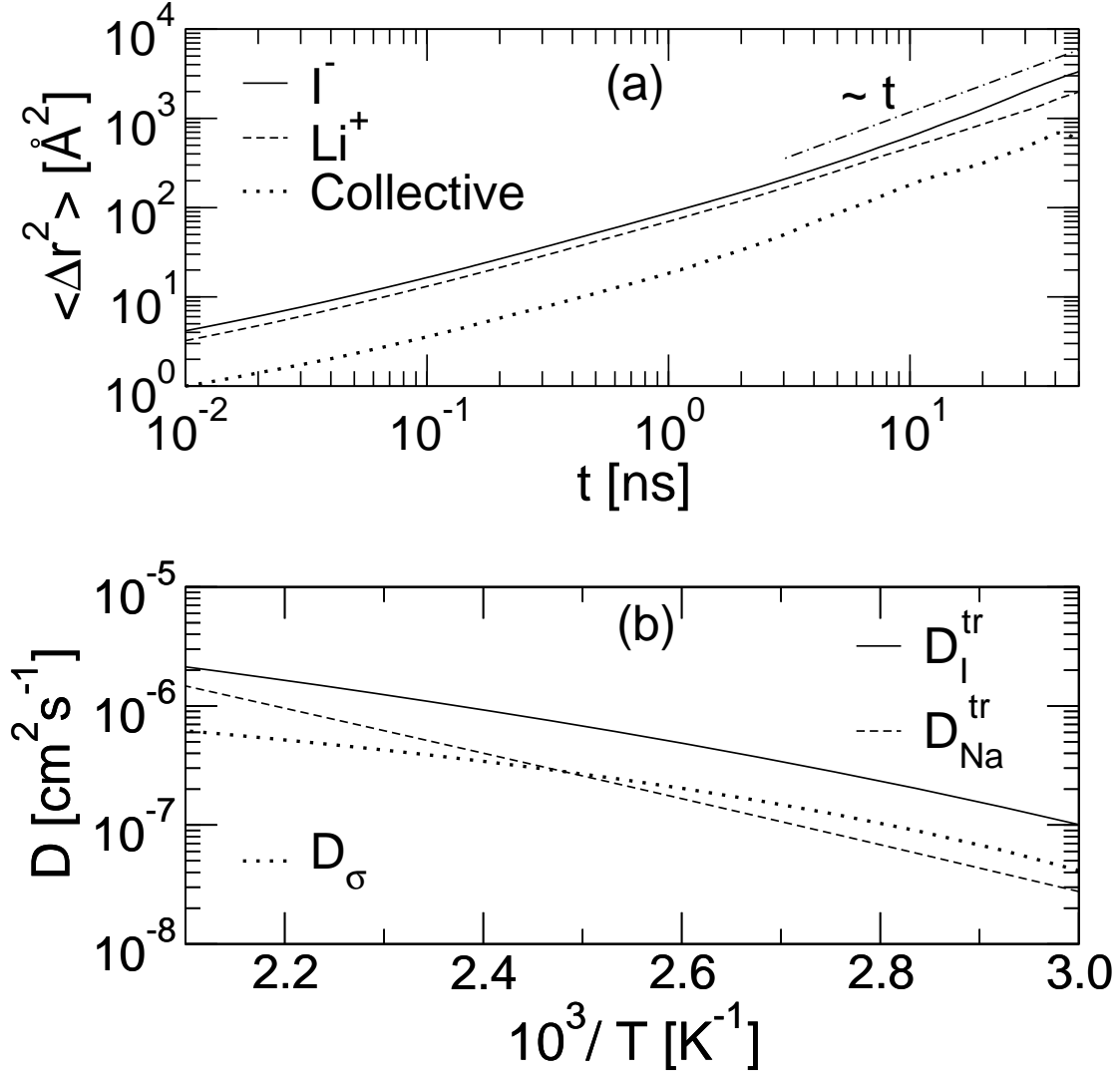}
  \caption{\label{DiffExptSim} (a) Mean square displacement of (MSD) Li$^+$, I$^-$ and collective displacement from a
  MD simulation of PEO/LiI (EO:Li$^+$ = 20:1) system at 450 K. The linear relationship between MSD and time is shown by a dot-dashed line. (b) Tracer diffusivities of
  Na$^+$ ($D_{Na}^{tr}$) and I$^-$ ($D_{I}^{tr}$) and collective diffusivity (from dc conductivity experiment)
  in PEO/NaI with EO:Na$^+$=30:1 from Ref.\ \cite{Stolwijk04}. Note that in Ref.\ \cite{Stolwijk04}  
  the collective diffusivity $D_\sigma^\prime$ was defined as $D_\sigma^\prime = 2D_\sigma$.
  }
  \vspace{5pt}
\end{figure}

The mean square displacement (MSD) of Li$^+$ and I$^-$ ions
obtained from the MD simulation  at T = 450 K are displayed in
Fig.\ \ref{DiffExptSim}(a). The I$^-$ ions are slightly more mobile (by a factor
of $\approx 1.6$) compared to the Li$^+$ ions. The collective
diplacement of the ions [see Eq.\ \eqref{kubo}] which measures
the mean square displacement of the center of charge of the system
is also shown in Fig.\ \ref{DiffExptSim}(a). The collective diffusivity,
$D_\sigma$ is diminished by a factor of approximately 3 with respect
to the average diffusivity of the ions, $(D_{Li} + D_{I})/2$.
Fig.\ \ref{DiffExptSim}(b) shows the experimental values of the tracer
diffusivities of Na$^+$ ($D_{Na}^{tr}$) and I$^-$ ($D_{I}^{tr}$)
in PEO/NaI (EO:Na=30:1 and polymer chain lengths were of the order
of $10^5$) from Ref.\ \cite{Stolwijk04} as a function of
temperature. The collective diffusivity for this system was
reckoned from dc conductivity using Eq.\ \eqref{nernst}. At T =
450 K factors of $D_I^{tr}/D_{Na}^{tr} \approx 1.7$ and
$(D_{Na}^{tr} + D_I^{tr})/2D_\sigma \approx 2.4$ are observed for
the experimental system similar to the simulated PEO/LiI system.
The absolute values of the diffusivities of the ions in the
simulated and the experimental system are of the same order of
magnitude at the considered temperatures.
Furthermore, the reduction of the ionic diffusivity [i.e.\ $0.5(D_I^{tr} + D_{Li}^{tr}) - D_\sigma$] is
close to the experimentally observed value.

From the simulation we identify a non-free ion as one which is associated with a
counterion such that it is separated by a distance of less than 4.6 \AA\
that demarcates the first minimum of the \hyp{Li}{I} radial
distribution function, $g_{LiI}$ [see Fig.\ \ref{corrfn} row 1, column
2]. Depending upon the number (one or two) of I$^-$ ions in the nearest neighbor shell of
a Li$^+$ ion one
may speak of an ion-pair or an ion-triplet, respectively. The
same cut-off radius is applied for defining a free or non-free I$^-$.
With this criterion the proportions of free and non-free ions present
in the simulated system are displayed in Tab.\ \ref{tabassoc}.

\begin{table}
\caption{\label{tabassoc} Proportion of free ($n=0$) and non-free ($n=1$: pairs, $n=2$:   triplets) ions in percentage.}
\begin{tabular}{| l | c |  c | c | c | c |} \hline 
     $n$ & 0 & 1 & 2  & 3 \\ \hline 
 Li [I]$_n$ & 6.2 & 75.9 & 17.9  & 0 \\ \hline
 I [Li]$_n$ & 9.3 & 69.8 & 20.6 & 0.3 \\  \hline
\end{tabular}
\end{table}

\begin{figure}
  \vspace{5pt}
  \includegraphics[clip,width=\linewidth]{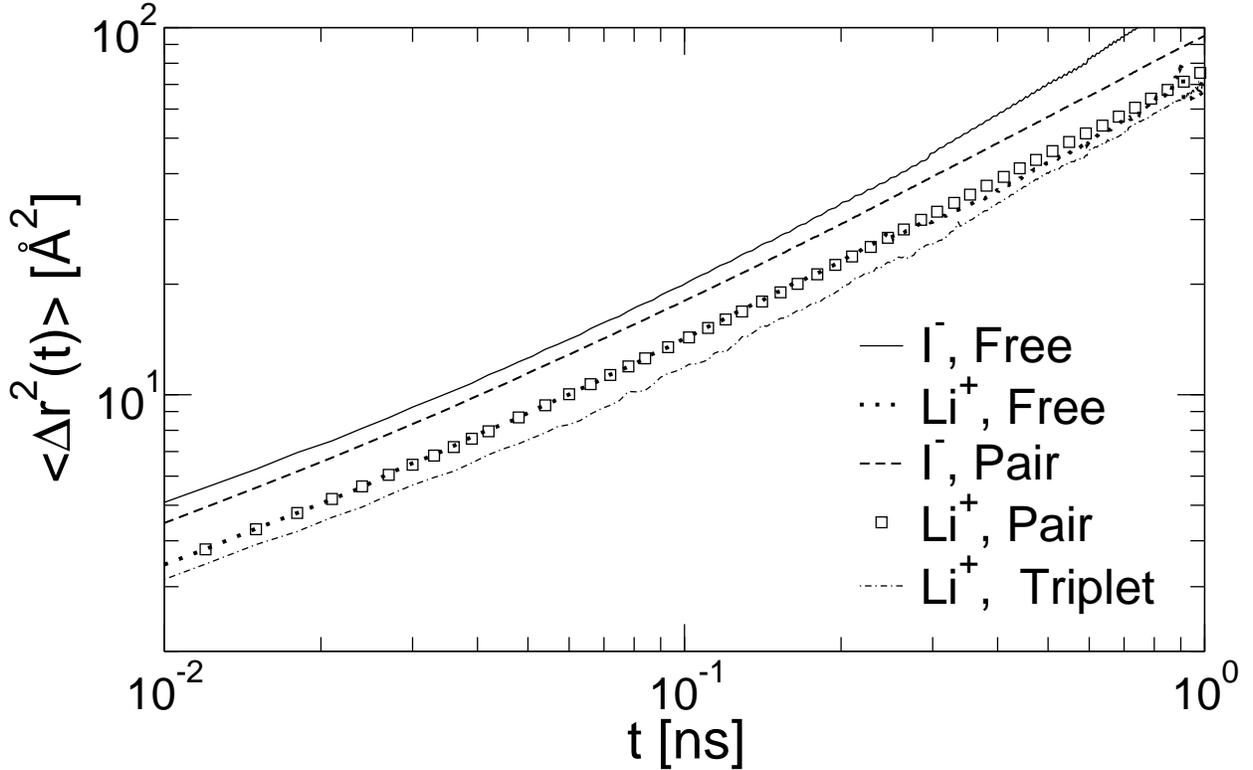}
  \caption{\label{MsdFrPairs}Mean square displacement of free I$^-$ (solid), free Li$^+$ (dashed), Li$^+$ with one I$^-$ within its radius of 4.6 \AA\ ($\square$) and Li$^+$ with two I$^-$ ions within its radius of 4.6 \AA\ (dot-dash).}
\
  \vspace{5pt}
\end{figure}

We have computed the MSD of a subset of
Li$^+$ or I$^-$ ions which are either free or participating in an
ion-pair or ion-triplet (see Fig.\ \ref{MsdFrPairs}).
An ion was considered free during
time $t$ if it had no counterion within its nearest neighbor distance
for at least $95 \%$ of the time $t$. Also shown are the MSD of those
Li$^+$ ions which were paired with only one I$^-$ (but not
necessarily with the same I$^-$) for $95 \%$ of the time. A
similar definition was used for calculating the MSD of Li$^+$
possessing two I$^-$ neighbors (i.e. ion-triplets). This definition
limits the maximum values for $t$ because of the finite life-time [see Ref.\ \cite{BoroLiIDynamic}]
of a specific local structure. The typical life-times of ion-pairs
will be discussed in more detail further below.

It is observed that the dynamics of the free Li$^+$ ions are
almost indistinguishable from the paired Li$^+$ ions in dramatic
contrast to the result of the data fitting, reported in
\cite{Stolwijk04}. The present result is expected since most of
the Li$^+$ ions in the simulated system are coordinated to three
to six contiguous ether oxygen atoms of the PEO backbone and all cations
are found to be complexed by EO atoms. Thus, the
Li$^+$ dynamics mainly reflects the dynamics of the polymer chain.
In particular, this strongly correlates the dynamics of free
Li$^+$ ions and paired Li$^+$ ions. The dynamics of the subset of
free I$^-$ are faster than the free Li$^+$ ions by a factor of
1.6. The MSD of Li$^+$ ions in the triplets are marginally slower
compared to the Li$^+$ in the pairs (Fig.\ \ref{MsdFrPairs}). The I$^-$ in the
ion-pairs are faster than the Li$^+$ in the ion-pairs
for short times ($t < 2$ ns) and beyond this merge together. The difference
at short times reflects the restriction of Li$^+$ motion due to the strong
coupling to the polymer.

\begin{figure}
 \includegraphics[clip,width=0.9\linewidth]{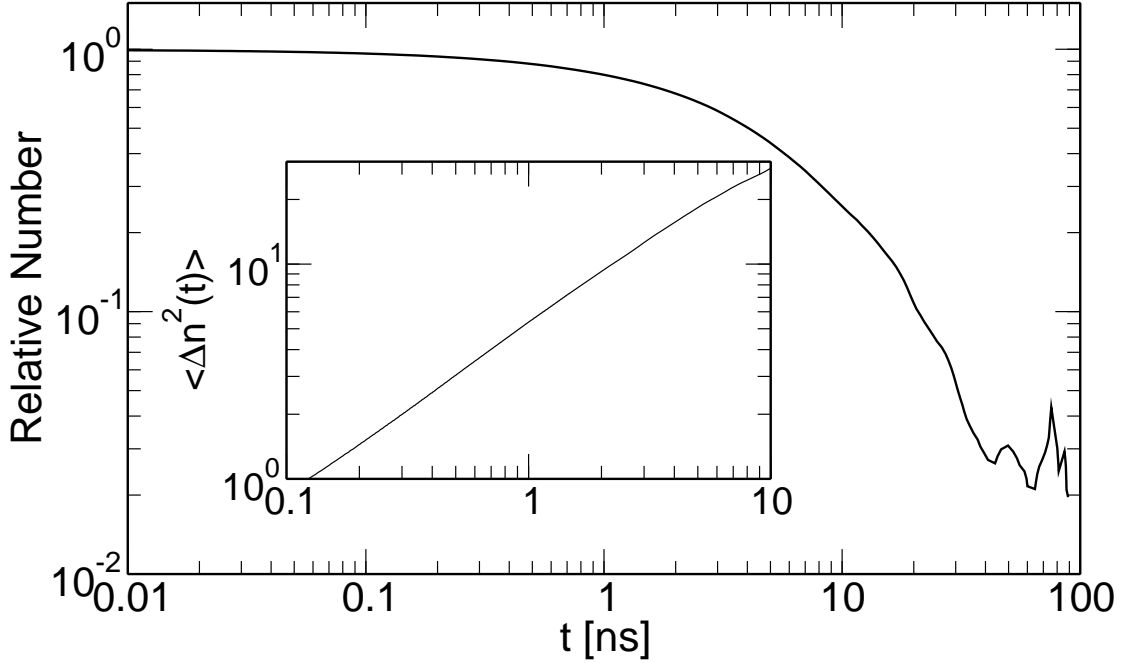}
 \caption{\label{probLiI} Relative likelihood that a lithium ion is complexed to a specific iodide ion at times $t=t_0$ and $t=t_0 + t$. [Inset] Mean squared variation in the average ether oxygen index, $\langle \Delta n^2(t) \rangle$, to which a Li$^+$ ion is coordinated while still being constrained to the same polymer chain in time $t$.}
\end{figure}

Even though the typical life-time of a \hyp{Li}{I} association only lasts for about
700 ps \cite{BoroLiIDynamic}, there exist forward and backward jumps of the ions leading to an extension of the time duration that an ion is complexed to the same counterion. To see this clearly, Fig.\ \ref{probLiI} shows the probability that a particular pair of Li$^+$ and I$^-$ separated by less than 4 \AA\ exists at times $t=t_0$ and $t=t_0 + t$ where $t_0$ is some reference time. On an average about 6 ns is required for a Li$^+$ ion to change its I$^-$ neighborhood.
Over this timescale a Li$^+$ ion translates along a polymer chain by a distance of approximately 4 monomers. Thus, the \hyp{Li$^+$}{I$^-$} pairs move collectively along a chain. This is seen from \figref{probLiI} (inset) where the mean square variation of the ether oxygen index $\langle \Delta n^2 (t)\rangle$ experienced by a Li$^+$ while being still constrained to the same chain is plotted in time $t$ (also see Ref.\ \cite{MHprl}). \\

\section{\bf Ionic Correlations}

The collective diffusivity can be calculated from the  microscopic
dynamics of the ions via
\begin{subequations}\label{kubo}
\begin{align}
D_\sigma & = \lim_{t \rightarrow \infty} \frac{1}{6 t n} \sum_{i,j=1}^{n} z_i z_j \langle \Delta\mathbf{R}_i(t) \Delta\mathbf{R}_j(t) \rangle \label{kubo1} \\
         & = \lim_{t \rightarrow \infty} \frac{1}{6tn} \Bigl [ \sum_{i=1}^{n} \langle \Delta \mathbf{R}_i^2(t) \rangle + \sum_{i,j \neq i}^{n} z_i z_j \langle \Delta \mathbf{R}_i(t) \Delta \mathbf{R}_j(t) \rangle \Bigr ]\label{kubo2}\\
& \equiv D_{avg} - \lim_{t \rightarrow \infty}
\frac{\mathcal{E}_{cross}(t)}{6 t}\label{kubo3}
\end{align}
\end{subequations}
where $\Delta \mathbf{R}_i(t) \equiv \mathbf{R}_i(t) -
\mathbf{R}_i(0)$ is  the displacement of a specific ion $i$
carrying a charge of $z_i$ during time $t$. Whereas the {\it self}
terms [i.e. the first term of Eq.\ \eqref{kubo2}] express the
diffusivities of all ions (i.e. $D_{avg}$),  the {\it cross}
terms $\mathcal{E}_{cross}(t)$ contain the {\it correlation}
effects between distinct ions and give rise to a non-trivial Haven
ratio.

\begin{figure*}
  \includegraphics[clip,width=0.7\linewidth]{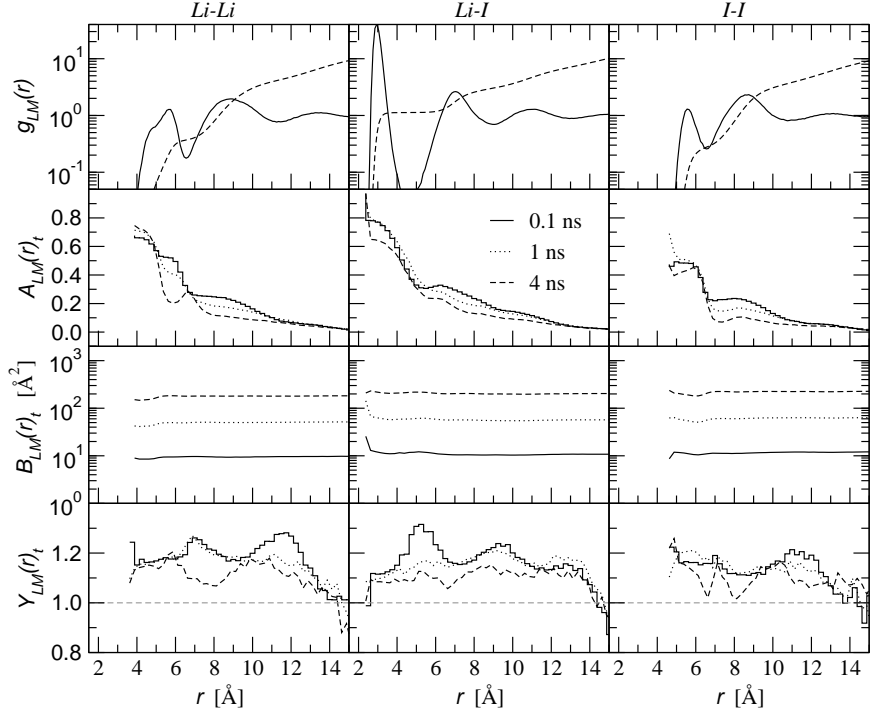}
  \vspace{5pt}
  \caption{\label{corrfn} Correlation functions: Each column displays a characteristic function of
  a specific pair of ions (e.g.$Li-Li$, $Li-I$, $I-I$) and each row displays the specific characteristic
  [e.g.\ radial distribution function or rdf, $g_{LM}(r)$; directional correlation functions $A_{LM}(r,t)$, mobility correlation
  function $B_{LM}(r,t)$ and the correction term $Y_{LM}(r,t)$ in Eq.\ \eqref{elmt}]. The first row also shows the
  integrated rdf, $\frac{\rho}{2} \int_0^r dr \: 4 \pi r^2 g_{LM}(r) $ [dashed line].}
\end{figure*}

The second term of Eq.\ \eqref{kubo3} can be rewritten in terms
of its different pair contributions. Using $z_{Li^+} = 1$ and
$z_{I^-} = -1$ one obtains
\begin{equation}\label{EX}
\mathcal{-E}_{cross}(t) = \mathcal{E}_{LiLi}(t) - \mathcal{E}_{LiI}(t) + \mathcal{E}_{II}(t)
\end{equation}
where
\begin{equation}\label{ELM}
\mathcal{E}_{LM}(t) = \frac{1}{n} \sum_i^{\{L\}} \sum_{j \ne i}^{\{M\}} \langle \Delta \mathbf{R}_i(t) \Delta \mathbf{R}_j(t) \rangle.
\end{equation}
Here, $\{L\}$ denotes a set of cations or anions and the condition
$j \ne i$ in the second summation of Eq.\ \eqref{ELM} implies that
only {\it distinct} pairs of ions are considered. The negative
sign in the middle term of the rhs of Eq.\ \eqref{EX} results
from the opposite signs of the charges on Li$^+$ and I$^-$ ions.
One expects in analogy to inorganic ion conductors a positive
correlation of all ionic species, i.e.  $\mathcal{E}_{LM}(t)
> 0$. However, since the correlation between  Li$^+$ and I$^-$ ions
will turn out to be particularly pronounced, for a polymer electrolyte,  one gets
$\mathcal{E}_{cross}(t)>0$ and $H > 1$ in contrast to inorganic
ion conductors where one only has the contribution of
cation-cation pairs.

Our goal is to elucidate the origin of the correlations, contributing
to $\mathcal{E}_{LM}(t)$. For this purpose we rewrite
\begin{equation}
\Delta \mathbf{R}_i(t,t_0) \Delta \mathbf{R}_j(t,t_0) = \\  \Delta
\widehat{\mathbf{R}}_i(t,t_0) \cdot  \Delta
\widehat{\mathbf{R}}_j(t,t_0) \: | \Delta \mathbf{R}_i(t,t_0) |
\cdot | \Delta \mathbf{R}_j(t,t_0) |
\end{equation}
where  $ \Delta \mathbf{R}_i(t,t_0) =  \mathbf{R}_i(t+t_0) -
\mathbf{R}_i(t_0)$ denotes the  displacement vector of ion $i$
from its initial position at time $t=t_0$ to its final position at
$t=t_0+t$ and $\Delta \widehat{\mathbf{R}}_i(t,t_0) =  \Delta
\mathbf{R}_i(t,t_0) / |\Delta \mathbf{R}_i(t,t_0)| $ refers to the
unit vector of the displacement during time $t$ of ion $i$
starting from its initial location at $t=t_0$. Thus, a necessary
condition for a non-zero $\mathcal{E}_{LM}(t)$ is the presence of
directional correlations as expressed by the scalar product of the respective unit
vectors. Qualitatively, one would expect that they mainly emerge
from nearest-neighbor ions. Furthermore, the absolute value of
$\mathcal{E}_{LM}(t)$ might further increase, if nearby pairs are
particularly mobile.

To quantify these different contributions to $\mathcal{E}_{LM}(t)$
we introduce the following correlation functions between two
distinct ions $i$ and $j$
\begin{subequations}\label{cde}
\begin{align}
a_{ij}(t;t_0) &=  \Delta \widehat{\mathbf{R}}_i(t,t_0) \cdot  \Delta \widehat{\mathbf{R}}_j(t,t_0) \\
b_{ij}(t;t_0) &=  | \Delta \mathbf{R}_i(t,t_0) | \cdot  | \Delta
\mathbf{R}_j(t,t_0) |.
\end{align}
\end{subequations}
The function $a_{ij}(t;t_0)$ represents the directional
correlation of two ions whereas $b_{ij}(t;t_0)$ reflects their
joint mobility during times $t=t_0$ and $t=t_0+t$.

In a first step these functions will be averaged over pairs of ions
separated by distance $r$. In general, the distance between two ions
will change with time.  A convenient way to take this into
account is to use the initial and final inter-ionic separation distance
with equal weight for the identification of $r$. With an additional average over
$t_0$ and using the abbrevation $r_{ij}(t) = |\mathbf{R}_i(t)
- \mathbf{R}_j(t) | $ one can therefore define
\begin{subequations}\label{avg}
\begin{align}
 \langle a_{ij} (r,t) \rangle  &= \frac{1}{2N_{ij}} \langle \{\delta_{r, r_{ij}(t_0)} + \delta_{r, r_{ij}(t_0 + t)}\} a_{ij}(t;t_0) \rangle_{t_0} \\
 \langle b_{ij} (r,t) \rangle  &= \frac{1}{2N_{ij}} \langle \{\delta_{r, r_{ij}(t_0)} + \delta_{r, r_{ij}(t_0 + t)}\} b_{ij}(t;t_0) \rangle_{t_0}
\end{align}
\end{subequations}
where $\langle \cdots \rangle_{t_0}$ denotes the average over the
time origins; $N_{ij} = \langle \delta_{r, r_{ij}(t_0)} +
\delta_{r, r_{ij}(t_0 + t)} \rangle_{t_0}$ is the normalization
constant [proportional to $\int dr r^2 g(r)$] and $\delta_{r,r_{ij}(t)}$ is
the Kronecker delta function which bins the quantities
$a_{ij}(t;t_0)$ and $b_{ij}(t;t_0)$ at both the initial and final
distance between ions $i$ and $j$.
Next, averaging the functions over all distinct pairs of ions, we
define
\begin{subequations}\label{ABdefn}
\begin{align}
A_{LM}(r,t) &= \langle a_{ij}(r,t) \rangle_{i \epsilon \{L\}; \: j \epsilon \{M\}}  \\
B_{LM}(r,t) &= \langle b_{ij}(r,t) \rangle_{i \epsilon \{L\}; \: j \epsilon \{M\}}.
\end{align}
\end{subequations}
In the last step one can average $A_{LM}(r,t)$ 
over all distances. This yields
\begin{equation}\label{IntegA}
  \mathcal{A}_{LM}(t) = (2 - \delta_{L,M})\int dr \, n_{LM}(r)  A_{LM}(r,t)
\end{equation}
where we have defined
\begin{equation}\label{nLM}
n_{LM}(r) = 4 \pi r^2 \: \frac{\rho}{2} \: g_{LM}(r)
\end{equation}
and $\rho = n/V$ is the number density of the ions (cations plus
anions).  The prefactor $2-\delta_{L,M}$ in Eq.\ \eqref{IntegA} takes into account
that the total number of unlike pairs ($L \neq M$) is twice the number of like ($L = M$) pair of ions
. In Eq.\ \eqref{nLM} the number density per species is expressed as $\rho/2$
and the radial distribution function (rdf) for the pair of species $L-M$ is referred to as $g_{LM}(r)$.

Likewise, one can define the quantity $E_{LM}(r,t) = \langle a_{ij}(r,t) b_{ij}(r,t) \rangle_{i \epsilon \{L\}; j \epsilon \{M\}}$ using relationships similar to Eq.\ \eqref{avg} and \eqref{ABdefn}.
Our goal is to express $\mathcal{E}_{LM}(t)$ (see Eq. \ref{ELM}) in terms of
more elementary contributions. Based on the definitions, introduced
so far, we can now write
\begin{equation}
\begin{split}
\label{elmt}  \mathcal{E}_{LM}(t) &= (2 - \delta_{L,M}) \int dr \, n_{LM}(r) \, E_{LM}(r,t) \\ &= (2 - \delta_{L,M}) \int dr \, n_{LM}(r) \,  Y_{LM}(r,t)
A_{LM}(r,t) B_{LM}(r,t) 
\end{split}
\end{equation}
which implicitly defines the function $Y_{LM}(r,t)$.
If the directional correlation [via $A_{LM}(r,t)$] and the mobility
correlation [via $B_{LM}(r,t)$] of a pair of ions are statistically
uncorrelated then one has $E_{LM}(r,t) = A_{LM}(r,t) B_{LM}(r,t)$,
i.e. $Y_{LM}(r,t)=1$. Possible correlations can be taken into
account by slightly different values of $Y_{LM}(r,t)$. Fig.\ \ref{corrfn}
depicts the functions $g_{LM}(r)$, $A_{LM}(r, t)$, $B_{LM}(r, t)$ and $Y_{LM}(r, t)$
for the different pairs of ionic species [$(L,M) \; \epsilon\ \{Li,
I\}$]. The correlation functions, $A_{LM}(r,t)$ and $B_{LM}(r,t)$
are evaluated at specific times $t$ = 0.1 ns, 1 ns and 4 ns.
One can see that $Y_{LM}(r, t)$ is larger than one. The resulting
correlation between the directional and the mobility correlation, however,
is rather weak.

The high proportion of ion-pairs (\hyp{Li}{I}) present in the system
manifests as a pronounced nearest neighbor peak in the $g_{LiI}(r)$
at $r=2.9$ \AA. In contrast, the rdfs between the like charges,
$g_{LiLi}(r)$ and $g_{II}(r)$ are of similar strength in their
respective first and second nearest neighborhood but exhibit weaker
local density when compared to $g_{LiI}(r)$. To quantify this effect
we introduce
\begin{equation}\label{nLMShell}
n_{LM}^{(k)}  \equiv \frac{\rho}{2} \int_{r_1}^{r_2} dr \: 4 \pi r^2 g_{LM}(r)
\end{equation}
as the number of particles of species $M$ in the $k$-th
neighborhood of a particles of species $L$ (Note that the
coordination numbers of $M$ around $L$ and $L$ around $M$, in the
different shells, are the same, considering equimolar amounts of
either ionic species). Here, $r=r_1$ and $r=r_2$ are the minima of
$g_{LM}(r)$ corresponding to the $k$-th neighborhood. The
values for $n_{LM}^{(k)}$ for $k=1,2,3$ are listed in Tab. \ref{tabLiIcoordn}.

\begin{table}
  \caption{\label{tabLiIcoordn} Coordination numbers in the first, second and third neighborhoods using Eq.\ \eqref{nLM}. The bounds $r_1$ and $r_2$ for each of the neighborhoods were determined from $g_{LM}(r)$ in the \figref{corrfn}.}
  \begin{tabular}{ | l | c | c | c | c |}\hline  
  \hyp{L}{M} &  $n^{(1)}_{LM}$ & $n^{(2)}_{LM}$ & $n^{(3)}_{LM}$ \tabularnewline[1mm] \hline 
  \hyp{Li}{Li}    &  0.4 & 3.8 & 5.5 \\ \hline
  \hyp{Li}{I}     &  1.1 & 1.9 & 3.7 \\ \hline
  \hyp{I}{I}      &  0.3 & 3.3 & 5.2 \\ \hline
  \end{tabular}
\end{table}

In case that the system would have {\it only} free ions and well-defined \hyp{Li}{I}
pairs the quantity $n_{Li I}^{(1)}$ would be identical to the
fraction $r_{nf}$ of non-free ions. However, about 20 \% of all
ions form ion-triplets. This increases the value of  $n_{Li
I}^{(1)}$ with respect to $r_{nf}$. Numerical analysis shows $r_{nf} \approx  0.92$
 (see Tab.\ \ref{tabassoc}) which is, indeed somewhat smaller than $n_{Li I}^{(1)}$.
The presence of these triplets also provides one
important contribution to $n_{Li Li}^{(1)} $ and $n_{I I}^{(1)} $
and thus to $ \mathcal{E}_{LiLi}(t)$ and $\mathcal{E}_{II}(t)$.
Due to the smallness of $n^{(1)}_{LiLi}$ and $n^{(1)}_{II}$
one already
anticipates that the contributions to $\mathcal{E}_{cross}(t)$ from the like
pairs is minor.

The joint mobility functions, i.e.\ the average of the product of
the length of the displacement vectors of two distinct ions during
time $t$, $B_{LM}(r,t)$, are plotted in the third row of \figref{corrfn}
for the species \hyp{Li}{Li}, \hyp{Li}{I} and \hyp{I}{I}. We see,
that this function is essentially independent of $r$ for all times
and is very similar for all the three types of pairs. In agreement
with \figref{MsdFrPairs} this means that spatial proximity of two ions does
not enhance its mobility.

In contrast, $A_{LM}(r,t)$ decreases with $r$ for all types of
pairs. Qualitatively, this is due to the screening of the
interactions between the ions.
Also, as time progresses
$A_{LM}(r,t)$ becomes smaller for fixed $r$. This is expected
because each ionic aggregate has a finite lifetime beyond which it
disintegrates. Each of the ions which were once part of an ionic
cluster will, beyond a certain time, dislodge, migrate and finally
participitate in the formation of another new cluster. We would like to mention that
 an ionic positional correlation function similar in spirit has been reported by
M\"{u}ller-Plathe et al \cite{Plathe2, Plathe}.

 \begin{figure*}
  \vspace{5pt}
  \includegraphics[clip,width=0.8\linewidth]{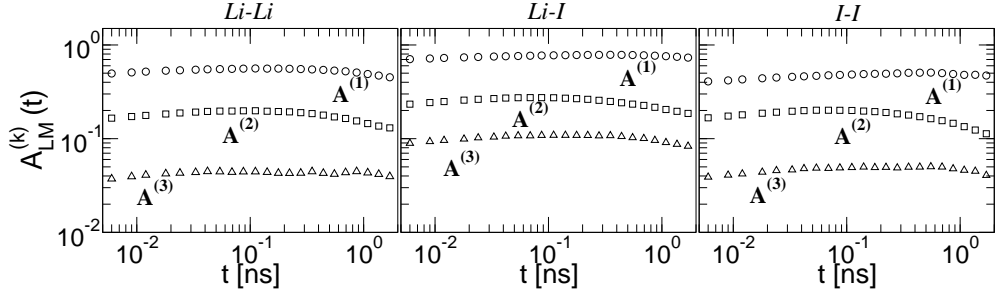}
  \caption{\label{ALMshell} Directional correlation contributions from the  $k$-th neighborhood of B-type ions
around A-type ions as a function of time, $A^{(k)}_{LM}(t)$, for
the distinct pairs of ions of  types $(A,B) \equiv$ $\{Li,I\}$.
Contributions from first ($k=1$; $\circ$), second ($k=2$;
${\scriptstyle \Box}$) and third ($k=3$; ${\scriptstyle
\triangle}$) neighborhoods are only displayed. }
  \vspace{5pt}
\end{figure*}

Among the general features of $A_{LM}(r,t)$ one can see that they
decay with $r$ in a step-like manner. The steps, representing
constancy in the correlation strength, are short spanned and
located at pair separation distances corresponding to the peak
positions of the $g_{LM}(r)$. Thus, it may seem reasonable to
characterize the average properties of the $A_{LM}(r,t)$ in the
individual shells.  For this purpose we define
\begin{equation}\label{ALM}
A_{LM}^{(k)}(t) = \frac{\int_{r_1}^{r_2} dr 4 \pi r^2 g_{LM}(r)
A_{LM}(r,t) }{ \int_{r_1}^{r_2} dr 4 \pi r^2 g_{LM}(r)}
\end{equation}
where $r=r_1$ and $r=r_2$ are the minima of $g_{LM}(r)$
corresponding to the $k$-th neighborhood  (see Fig.\ \ref{corrfn}).
Analogously one can define $B_{LM}^{(k)}(t)$ and $E_{LM}^{(k)}(t)$. 
In the
following we analyze the first three shells $A^{(k)}_{LM}$
$(k=1,2,3)$. Their time dependence is shown in Fig.\ \ref{ALMshell}. The
functions are more or less constant at short times.
However, these curves display a trend of decay with increasing time,
as expected.
Correlations for all pairs of
ionic species show the order: $A_{LM}^{(1)}(t) > A_{LM}^{(2)}(t)
 > A_{LM}^{(3)}(t)$ where $(L,M) \; \varepsilon \; \{Li,
I\}$. Henceforth, we will often refer to a timescale $t^\star = 0.5$ ns.
Note that $t^*$ is significantly smaller than the typical life-time of a
\hyp{Li}{I} pair.
For $t = t^\star$ we have $A^{(1)}_{Li
I}(t^\star) = 0.8$, $A^{(1)}_{Li Li}(t^\star) = 0.53$ and $A^{(1)}_{I
I}(t^\star) = 0.49$.  Thus, the directional strength of motion of
a pair of unlike charges (\hyp{Li}{I}) is more potent in comparison to
those of the like charges and is close to a perfect correlation $A^{(1)}_{Li
I}(t^\star) = 1$. Furthermore, the shell-dependence is due to the
screening of the higher shells of ions with respect to the central
ion.

After having discussed the individual contributions to
$\mathcal{E}_{LM}(t)$ we now come back to Eq.\ \eqref{ELM} by
rewriting it as
\begin{equation}\label{E12_2}
\mathcal{E}_{LM}(t)
  = (2- \delta_{L,M}) \sum_k  n^{(k)}_{LM} E^{(k)}_{LM}(t).
\end{equation}
The correlations originating from the $k$-th neighborhood
is contained in the quantity
$E^{(k)}_{LM}(t)$.
From the previous discussion we have seen that the dependence of $E_{LM}^{(k)}(t)$
on the ionic pair \hyp{L}{M} is mainly governed by
$A_{LM}^{(k)}(t)$ which for \hyp{Li}{I} is approximately 60 \% larger
than for the like pairs. Since, also, $n_{LM}^{(1)}$ is much larger
for \hyp{Li}{I} (see Tab.\ \ref{tabLiIcoordn}) the by-far dominant contribution
to $\mathcal{E}_{cross}(t)$ stems from $\mathcal{E}_{LiI}(t)$. \\

\section{\bf Stolwijk-Obeidi Model}

\begin{figure}
  \vspace{5pt}
  \includegraphics[clip,width=0.8\linewidth]{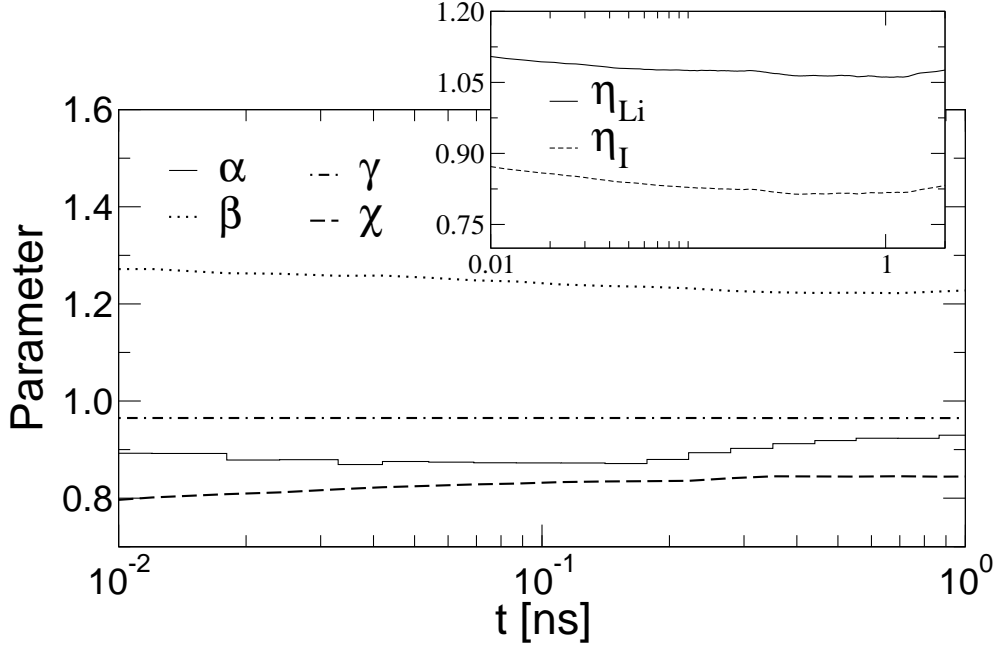}
  \caption{\label{params} Parameters $\alpha$, $\beta$, $\gamma$, $\chi$\:; 
and $\eta_{Li}$, $\eta_{I}$ (inset) measured as a function of time.}
  \vspace{5pt}
\end{figure}

In this section we begin with exact microscopic  expressions and
proceed through steps of simplifications to construct an
exact model similar in spirit to the phenomenological
Stolwijk-Obeidi (SO) model. 
The specific procedure is determined by the input from simulations.
The key idea of the SO model is to
express the observed single-particles diffusivities, obtained via
tracer diffusion, and the conductivity in terms of the properties
of single ions and ion-pairs. In the SO model no triplets are
taken into account. In case that these aggregates exist one should
therefore better speak in terms of {\it free} ions or {\it non-free}
ions, the latter possessing one or more counterion(s) in their
coordination sphere, i.e. existing as both ion-pairs
and ion-triplets in our case. In any event, this terminology gives
rise to a more general applicability of that approach.

First, we start to derive the single-particle dynamics. Here we
have to take into account that during the life-time of a pair the
MSD does not reach the linear regime. 
This requires a definition for the long time behavior of the transient
ionic states.
Specifically, we define the MSD  {\it associated} with ions of type $L$ (i.e.
either Li$^+$ or I$^-$) existing in state $S$ [i.e. either free
($f$) or non-free ($nf$)] in the following manner:
\begin{equation}\label{specMSD}
\langle \Delta R^2_{L,S} (t) \rangle = \frac{\sum_{i \varepsilon \{L\}}
 \langle [\delta_{s(i, t_0), S} + \delta_{s(i, t_0 +t), S}] 
\Delta \mathbf{R}^2_i(t,t_0) \rangle_{t_0}}{\sum_{i \varepsilon \{L\}} \langle 
\delta_{s(i, t_0), S} + \delta_{s(i, t_0+t), S } \rangle_{t_0}}.
\end{equation}
Here $s(i, t)$ stands for the instantaneous state of the ion $i$,
$\Delta \mathbf{R}^2_i(t,t_0)$ refers to the square of the
displacement experienced by ion $i$ between times $t=t_0$ and
$t=t_0+t$ and the Kronecker delta function $\delta_{s(i, t_0),S}$
picks the instantaneous state $S$ of ion $i$ at time $t_0$. For
$t \lesssim t^\star$ smaller than the typical timescale of the lifetime of nonfree ions
this definition would give rise to
the data in Fig.\ \ref{MsdFrPairs}. For $t \gg t^\star$, $\langle \Delta R^2_{L,f}(t) \rangle$ 
and $\langle \Delta R^2_{L,nf}(t)\rangle$ start to merge because of averaging effects, i.e.\
 free ions becoming non-free and vice versa.

The fraction of free or non-free ions can be written as
\begin{equation}\label{specr1}
r_{L,S} = \frac {\sum_{i \varepsilon \{L\}} \langle \delta_{s(i, t_0), S} +
\delta_{s(i, t_0+t), S} \rangle_{t_0} }{\sum_{i \varepsilon \{L\}} \langle
\sum_S [\delta_{s(i, t_0), S} + \delta_{s(i, t_0+t), S}]
\rangle_{t_0}}.
\end{equation}
Using this somewhat complex but generic definition one can show that Eq.\
\eqref{specMSD} and \eqref{specr1} can be combined (see Appendix) to yield
\begin{equation}\label{MSDion}
\langle \Delta R^2_L(t) \rangle = r_{L,f} \langle \Delta
R^2_{L,f}(t) \rangle + r_{L,nf} \langle \Delta R^2_{L,nf}(t)
\rangle.
\end{equation}
Note that the motion of state specific MSD is only given for times smaller than 
the average lifetime of a pair of ions Li$-$I and thus in particular we choose
$t = t^\star$.

The average dynamics of a non-free ion is characterised by
\begin{equation}\label{nfdyn}
\langle \Delta R^2_{nf}(t) \rangle = \frac{[r_{Li,nf}\langle \Delta R^2_{Li,nf}(t)
\rangle + r_{I,nf}\langle \Delta R^2_{I,nf}(t) \rangle]}{r_{Li,nf}
+ r_{I,nf}}.
\end{equation}
This definition can be used to relate the decrease in average
ionic diffusivity to the correlated motion of the ions. 
Eq.\ \eqref{EX} can be
rewritten via the following steps at a chosen timescale t$^\star$:
\begin{subequations}\label{chi15}
\begin{align}
\mathcal{E}_{cross}(t^\star) &=  \chi_1 \; \mathcal{E}_{LiI}(t^\star) & \\
         &=  \chi_1 \chi_2 \;  n^{(1)} E^{(1)}_{LiI}(t^\star) & \mbox{[cf. Eq.\ \eqref{E12_2}]} \\
         &=  \chi_1 \chi_2 \chi_3 \;  n^{(1)} A^{(1)}_{LiI}(t^\star) B^{(1)}_{LiI}(t^\star) &  \\
         &=  \chi_1 \chi_2 \chi_3 \chi_4 \; n^{(1)} B_{LiI}^{(1)}(t^\star)  &  \\
         &=  \chi_1 \chi_2 \chi_3 \chi_4 \chi_5 \; n^{(1)} \langle \Delta R^2_{nf}(t^\star) \rangle & \mbox{[cf. Eq.\ \eqref{nfdyn}]} \\
         &=  \chi_1 \chi_2 \chi_3 \chi_4 \chi_5 \chi_6\; r_{nf} \langle \Delta R^2_{nf}(t^\star) \rangle. &
         \label{chi15_e}
\end{align}
\end{subequations}
In the last step we have defined
\begin{equation}
r_{nf} = \frac{r_{Li,nf}+r_{I,nf}}{2}.
\end{equation}
The parameters $\chi_1, \chi_2,
\chi_3, \chi_4$, $\chi_5$ and $\chi_6$ are the dimensionless correction
factors required at every step of simplification in Eq.\
\eqref{chi15}. $\chi_1$ captures the cross correlations between
the like charges. Because the missing contributions to $\mathcal{E}_{cross}(t^\star)$
have opposite sign one naturally has $\chi_1 < 1$.  $\chi_2$ accounts for ignoring correlations
between Li$^+$ and I$^-$ ions beyond the nearest neighborhood
($\approx 4.6$ \AA).
One has $\chi_2 < 1$ because of the presence of anticorrelations between
the pair of ions beyond \mbox{$r > 16$ \AA}. The origin of the anticorrelations
is not yet comprehended fully, however, physically this might arise
from the tendency of the subensemble of particles to
nullify its momentum due to a finite simulation box size. 
$\chi_3$ is the correction due to the $Y$-factor [see Eq.\ \eqref{elmt}],
 which is given by the average value of $Y_{LM}(r,t)$ in the nearest neighbor shell,
reflecting correlations between the directional and the mobility correlations.
$\chi_4$  expresses the non-ideality of directional
correlation in the nearest-neighbor shell, i.e. $\chi_4 \approx A_{LiI}^{(1)}(t^\star)
\le 1$.
The factor $\chi_5$ is needed to approximate $B_{LiI}^{(1)}(t)$
by $\Delta R^2_{nf}(t)$. If one compares the quantities
$\Delta r_i \Delta r_j$ and $(\Delta r_i^2 + \Delta r_j^2)/2$
the latter is either greater than or equal to the former.
This rationalizes $\chi_5 < 1$.
 Lastly in the final step the factor
$\chi_6=n^{(1)}/r_{nf}$ is required which as discussed above is
larger than one. The individual values of
$\chi$ for $t = t^\star$ are listed in Tab.\ \ref{tabparams}.
\begin{table}
\caption{\label{tabparams}Parameters $\alpha$, $\beta$, $\gamma$, $\chi_i$, $\eta_{Li}$ and $\eta_I$ from simulation evaluated at time $t=t^\star=0.5$.}
\begin{tabular}{| c | c | c | c |c| c| c| c| c| c| c |} \hline 
 $\alpha(t^\star)$ & $\beta(t^\star)$ & $\gamma(t^\star)$ & $\chi_1(t^\star)$ & $\chi_2(t^\star)$ & $\chi_3(t^\star)$ & $\chi_4(t^\star)$ & $\chi_5(t^\star)$ & $\chi_6(t^\star)$ & $\eta_{Li}(t^\star)$ & $\eta_I(t^\star)$ \tabularnewline[1mm] \hline 
 0.959  & 1.221  &  0.966  & 0.94 & 0.88 & 1.1 & 0.86 & 0.92 & 1.18 & 1.07 & 0.82   \\ \hline
\end{tabular}
\end{table}

Next we define the following parameters for later convenience:
\begin{equation}\label{gambetalp}
\gamma \equiv \frac{r_{I,nf}}{r_{Li,nf}} ; \; \beta(t) = \frac{\langle \Delta R_{I, nf}^2(t) \rangle}{\langle \Delta
 R_{Li,nf}^2(t)\rangle}; \; \alpha(t) = \frac{\langle \Delta R_{Li,nf}^2(t) \rangle}{\langle \Delta R_{Li,f}^2(t
) \rangle}. 
\end{equation}
The time dependence of the
parameters $\alpha$, $\beta$ and $\gamma$ is shown in Fig.\ \ref{params}.
These are observed to be roughly stationary within the
1 ns time window which is less than the maximum life-time of an ion-pair. 
The values measured at $t=t^\star=0.5$ ns for the present system at T = 450 K can
be found in Tab.\ \ref{tabparams}. 
We have verified  
that choosing $t^\star = 0.1$ ns instead for defining the parameters would virtually yield similar 
magnitudes of results that are presented later.


That the value of $\alpha(t^\star$)
is less than unity (0.96 and 0.91 for T = 450 K and 425 K respectively), can be understood from the slightly slower dynamics of the
ion triplets than the free and the paired ions. It must be noted that $\alpha(t) \approx 1$ is a consequence of the coupling
between ion and segmental dynamics of PEO. On the other hand $\beta(t^\star)$ $\approx 1.2$
means that the I$^-$ ions in the pairs are slightly faster than their Li$^+$ counterparts
at short timescales (see also Fig.\ \ref{MsdFrPairs}).
It is interesting that the correction factor $\chi(t^\star)$ is found to be
less than unity (see Fig.\ \ref{params}) 
unlike the assumption of $\chi(t) = 1$
in the SO model.
In essence, all the fixed parameters
($\alpha(t^\star)$, $\beta(t^\star)$, $\gamma(t^\star)$ and $\chi(t^\star)$) are close to unity due to the
natural outcome of the ionic association and dynamics in the system and
which is why the SO model could approximately define these characteristics.
All these parameters are found to be nearly the same when going from the temperature of 450 K to 425 K (not shown). Even though $\chi_1(t^\star) \cdots \chi_6(t^\star)$ differs slightly between the temperatures, the final product is found out to be similar due to compensation of the increase of one factor by the decrease of another factor and vice versa.

The diffusion
constant $D_L$ can be written as $D_L = \lim_{\tau \rightarrow
\infty} \langle \Delta R^2_L(\tau) \rangle /6\tau$.
Furthermore, we define
\begin{equation}\label{DLs}
D_{L, S} = \lim_{t \rightarrow \infty} \frac{\langle \Delta R^2_L(t) \rangle}{\langle \Delta R^2_L(t^\star) \rangle}
 \frac{\langle \Delta R^2_{L, S}(t^\star)\rangle}{6t} = D_L \frac{\langle \Delta R_{L,S}^2(t^\star) \rangle}{\langle
 \Delta R_L^2(t^\star) \rangle}
\end{equation}
through which we are extrapolating the short time dynamics of the transient ionic state $S$ of species $L$ to long times
using the time dependence of $\langle R_L^2(t)\rangle$ as a reference. This extrapolation is necessary because $\langle R_{L,S}^2(t)\rangle$
is only defined for times of the order of the lifetime of the ion in state $S$ 
which is smaller than the timescale over which
the ions become completely diffusive. As a specific advantage of the definition of $D_{L,S}$, Eq.\ \eqref{MSDion} can now be rewritten as
\begin{equation} \label{SO}
D_{L}^{tr} = r_{L,f}D_{L,f} + r_{L,nf} D_{L,nf}.
\end{equation}
which is the form used in the SO model with $L \in \{ Li, I\}$.

In the SO model all $\chi_i$ are assumed to be one. According to
the above discussion this corresponds to the case of ideal \hyp{Li}{I}
pairs and no directional correlations for other pairs (either
pairs of like ions or pairs of Li$^+$ and I$^-$ ions beyond the nearest neighbor
shell).
Multiplying Eq.\ \eqref{chi15_e} with the factor $\mathcal{E}_{cross}(t)/6t\mathcal{E}_{cross}(t^\star)$ and using
 $D_{cross} \equiv \lim_{t \rightarrow \infty}
\mathcal{E}_{cross}(t)/6t$ one obtains from Eq.\ \eqref{chi15_e} and \eqref{nfdyn}
\begin{equation}\label{SO3}
2D_{cross} = \chi \; r_{nf} \left(\frac{2}{1 + \gamma} \eta_{Li} D_{Li, nf} + \frac{2 \gamma}{1 + \gamma} \eta_{I} D_{I, nf} \right).
\end{equation}
Here, $\chi$ is the product of the parameters
$\chi_1 \cdots \chi_6$. We have introduced the parameters $\eta_{Li}$ and $\eta_I$ which are required
for an exact relation between the diffusion coefficients in Eq.\ \eqref{SO3} and are defined 
as $\eta_{Li} \equiv \eta_{Li}(t^\star)$ and $\eta_I \equiv \eta_I(t^\star)$ with 
\begin{equation}\label{eta}
 \eta_L(t) = \frac{D_{cross}}{D_L}.\frac{\langle \Delta R_L^2(t) 
\rangle}{\langle \Delta R_{cross}^2(t) \rangle
 }.
\end{equation}
This parameter $\eta_L(t)$ (see inset of Fig.\ \ref{params}) captures the difference in timescales over which 
$\langle \Delta R^2_{cross}(t) \rangle$ and $\langle \Delta R^2_{L}(t) \rangle$ approach the linear regime, i.e.\
 the respective ionic dynamics become diffusive.
In the present case one has
$\eta_{Li} = 1.07$ and $\eta_I = 0.82$ by choosing $t^\star = 0.5 $ ns at T = 450 K (see Fig.\ \ref{params}).
$\eta_{Li} > \eta_I$ means that Li$^+$ becomes diffusive later than I$^-$ ions which
is evident from Fig.\ \ref{DiffExptSim}(a). As discussed in Ref.\ \cite{MHprl} the diffusive behavior of the Li$^+$ ions for long chains 
is achieved via jumps between different PEO chains.  
The residence time of the cation with one chain is found to be $\tau_{Li} \approx$ 100 ns \cite{MHprl}. 
Only for short chains (as in the present case) the timescale of Li$^+$ diffusion will be somewhat shorter because
of the dominance of the center of mass diffusion of the polymer. 
In contrast, an I$^-$ ion change its cationic neighbor(s) quite frequently over a timescale of $\tau_{I} \approx$ 7 ns
and therefore is able to become diffusive faster than its cationic counterparts.

In the following, we formally define $\alpha \equiv \alpha(t^\star)$, $\beta \equiv \beta(t^\star)$, $\chi \equiv \chi(t^\star)$. 
 The set of three equations, [Eq.\ \eqref{SO} with $L \: \varepsilon \: \{Li, I\}$ and
 Eq.\ \eqref{SO3}] is identical to the SO model if
$\chi=1$, $\gamma=1$,
$\beta = 1$ and $\eta_{Li} = \eta_I = 1$ 
i.e.\ if the picture of
strict \hyp{Li}{I} pairs would hold. However, these three equations
contain four parameters ($D_{Li, f}$, $D_{Li, nf}$, $D_{I, f}$,
$r_{nf}$).
Due to this ambiguity there are many
possible solutions for a given set of parameters. The solution,
presented in Ref.\ \cite{Stolwijk04} involved additional assumptions
about a similar temperature dependence of the parameters.
The resulting solution was $D_{Li, f} / D_{Li, nf} \approx
1/74$ for high temperatures.
To circumvent this problem we have introduced the dimensionless
parameter $\alpha(t)$ as in Eq.\ \eqref{gambetalp} and determined
its value (similar to $\beta(t)$, $\gamma(t)$, $\chi_i(t)$) from simulations. The reason for the observed 
magnitude of $\alpha(t)$ has been explained above. 

Using these parameters the relations of the generalized SO-model (see also Appendix II)
can be solved: 
\begin{subequations}\label{SOmod}
\begin{gather}
r_{nf} = \frac{1 + \gamma}{\chi \alpha \eta_{Li} (1 + \beta \gamma) D_{Li}^{tr}/D_{cross} + 2(1 - \alpha)} \\
D_{Li, f} = \frac{D_{Li}^{tr}}{1 - 2 r_{nf} (1 - \alpha)/(1 + \gamma)}
\\
D_{I,f}  = 
(D_I^{tr} - \frac{2 \alpha \beta \gamma \eta_{Li} r_{nf} D_{Li, f}}{\eta_I(1+ \gamma)})/( 1 - \frac{2 r_{nf} \gamma}{1 + \gamma}).
\end{gather}
\end{subequations}
The quantities $D_{Li}^{tr}$, $D_{I}^{tr}$ and $D_{cross}$ are
experimentally accessible.
The parameters $\alpha$,
$\beta$, $\gamma$, $\chi$, $\eta_{Li}$ and $\eta_I$ can be gathered from simulations,
thereby permitting a solution for the unknowns $r_{nf}$,
$D_{Li,f}$ and $D_{I,f}$ from Eq.\ \eqref{SOmod}. We neglect
the possible temperature dependence of $\alpha$, $\beta$, $\gamma$,
$\chi$ and $\eta_L$ because it is difficult to find a priori arguments
to estimate the temperature dependence of the four parameters.

\section{\bf Application of Stolwijk-Obeidi Model}

First, we aply the SO model to our simulation data.
We employ the parameters of Tab.\ \ref{tabparams} and 
$D_{Li}^{tr} = 6.7 \: \mathsf{X} \: 10^{-7}$ cm$^2$s$^{-1}$, $D_{I}^{tr} = 10.95 \: \mathsf{X} \: 10^{-7}$ cm$^2$s$^{-1}$
and $D_{\sigma}^{tr} = 2.7 \: \mathsf{X} \: 10^{-7}$ cm$^2$s$^{-1}$ (obtained through Fig.\ \ref{DiffExptSim}(a) from the 
simulation at T = 450 K) 
that gives $D_{cross} = 2.125 \: \mathsf{X} \: 10^{-7}$ cm$^2$s$^{-1}$. These when plugged into Eq.\ \eqref{SOmod} yields
the following: $r_{nf} = 0.915$, $D_{Li,f} = 6.966 \: \mathsf{X} \: 10^{-7}$ cm$^2$s$^{-1}$ and 
$D_{I,f} = 13.865 \: \mathsf{X} \: 10^{-7}$ cm$^2$s$^{-1}$ and thus $D_{I,f}/D_{Li,f}=2$. 
From Eq.\ \eqref{eta} one should then obtain $\frac{\langle \Delta R_{I}^2(t^\star) \rangle}{\langle \Delta R_{Li}^2(t^\star) \rangle} = \frac{D_{I,f}}{D_{Li,f}}\frac{\eta_{I}}{\eta_{Li}} \approx 1.5$. These, when compared to the actual  
values culled directly from the simulation, e.g.,  
$r_{nf}=0.92$ (see Tab.\ \ref{tabassoc}) and 
$\langle \Delta R_{I,f}^2(t^\star) \rangle/\langle \Delta R_{Li,f}^2(t^\star) \rangle=1.6$ (from Fig.\ \ref{MsdFrPairs}) are reproduced,
in good agreement
with the solutions of Eq.\ \eqref{SOmod}. By letting $\eta_{Li}=\eta_I=1$ and the other parameters of Tab.\ \ref{tabparams}
unchanged one obtains $r_{nf}=0.97$ and $D_{I,f}/D_{Li,f} \approx 10$ from Eq.\ \eqref{SOmod}. Clearly, the factors $\eta_{Li}$ and $\eta_I$ are
important.

Next, we apply the model on the same representative PEO/NaI polymer
electrolyte with the concentration of EO:Na$^+$ = 30:1 as Stolwijk and Obeidi
did. Accordingly, we use the tracer diffusivities of Na$^+$
($D_{Na}^{tr}$) and I$^-$ ($D_{I}^{tr}$) and the collective
diffusivity ($D_{\sigma}$) from Ref. \cite{Stolwijk04} at
different temperatures as input. Note the experimental data
that have been used subsequently in this work is calculated back
analytically from the fit results of Ref.\ \cite{Stolwijk04} using the VTF temperature
dependence of the single ion
diffusivity prefactors and the Arrhenius temperature dependence of
the pair formation constant (see details in Ref.\ \cite{Stolwijk04}).
Four different sets for the parameters $\alpha
$, $\beta$, $\gamma$, $\chi$, $\eta_{Na}$ and $\eta_I$ are considered which 
we will refer, henceforth, as SET I to SET IV and tabulated in Tab.\ \ref{tabset}.
This will allow delineation of the relative importance of the parameters.
\begin{table}
\caption{\label{tabset}Sets of parameters $\alpha$, $\beta$, $\gamma$, $\chi$, $\eta_{Li}$ and $\eta_I$ used for solving Eq.\ \eqref{SOmod}.}
\begin{tabular}{| l | c | c | c | c |c| c |} \hline 
 SET & $\alpha$ & $\beta$ & $\gamma$ & $\chi$ & $\eta_{Na}$ & $\eta_I$ \tabularnewline[1mm] \hline 
 SET I   & 0.959  & 1.221  &  0.966  & 0.848 & 1.065 & 0.818   \\ \hline
 SET II  & 0.959  & 1.221  &  0.966  & 0.848 & 1.0   & 1.0     \\ \hline
 SET III & 0.959  & 1.0    &  1.0    & 1.0   & 1.0   & 1.0     \\ \hline
 SET IV  & 74     & 1.0    &  1.0    & 1.0   & 1.0   & 1.0     \\ \hline
\end{tabular}
\end{table}
SET I has been retrieved from our simulation (same as in Tab.\ \ref{tabparams}) and additionally, we assume
this to be similar to the case of Na$^+$ ion in
PEO. SET IV corresponds to the original SO model of Ref.\ \cite{Stolwijk04}. Furthermore, we have superimposed 
a weak noise having a strength
between $\pm 5 \%$ to the experimental data [i.e $D_{Na}^{tr}(T)$,
$D_{I}^{tr}(T)$ and $D_{\sigma}(T)$] in order to evoke an impression
about the stability of the solution ($r_{nf}$, $D_{Na, f}$ and
$D_{I, f}$). The mean and standard
deviation of the unknowns are calculated for each temperature
from 100 independently noised realisations. We had also attempted mixing $\pm 10 \%$ noise
to the experimental data, however, the standard deviations of the solution exceeded mean values
in the high temperature limit. Thus, the accuracy of the input experimental data is imperative
in deriving the estimates of the solutions. 
   
\begin{figure}
  \includegraphics[clip,width=0.9\linewidth]{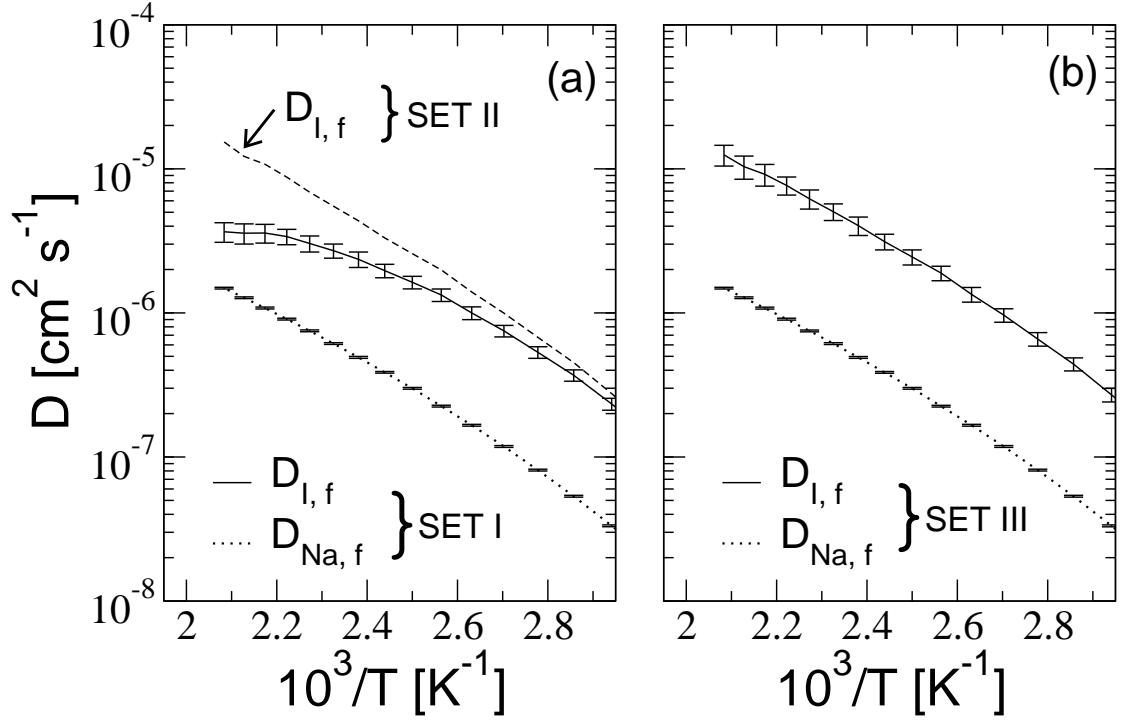}
  \caption{\label{DiffSO} Diffusivities of free I$^-$ (solid line) and Na$^+$ (dotted line) obtained from Eq.\ \eqref{SOmod}. Experimental data used is back-calculated from the final results in Ref.\ \cite{Stolwijk04} and then superimposed with a weak noise of $\pm 5$ \%. The standard deviation is shown by the vertical bars. Parameters used for the different sets are listed under Tab.\ \ref{tabset}.}
\end{figure}

\begin{figure}
  \includegraphics[clip,width=0.9\linewidth]{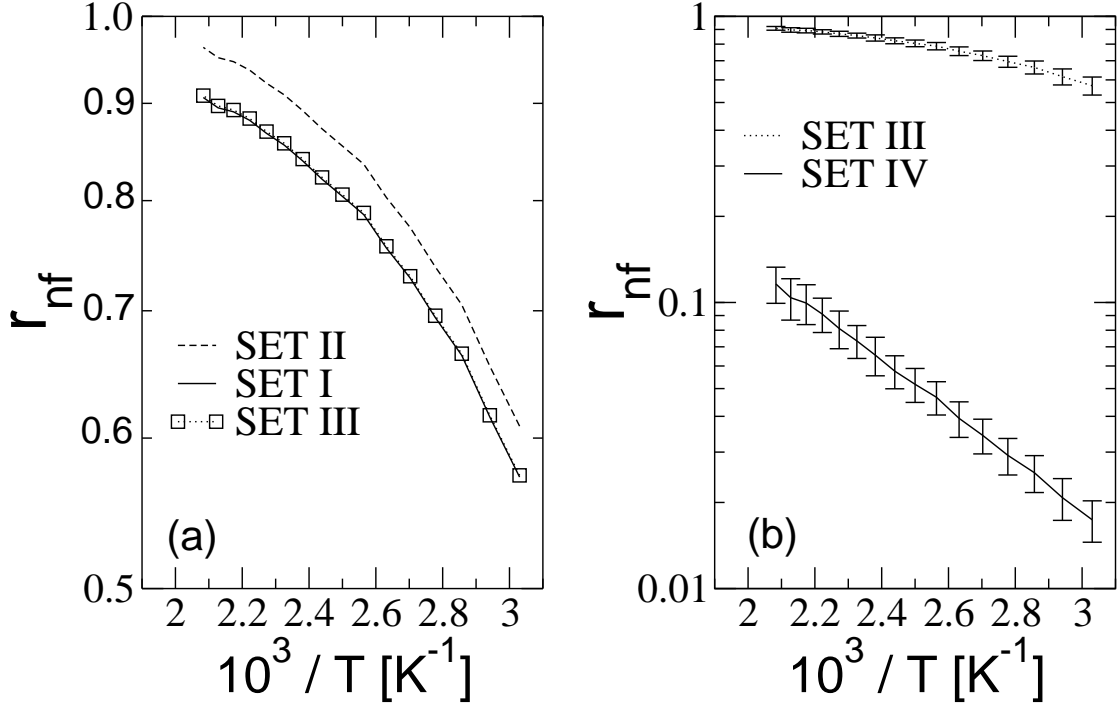}
  \caption{\label{FrNonFreeSO} Fraction of non-free ions ($r_{nf}$) versus inverse temperature (T) obtained from Eq.\ \eqref{SOmod}. Experimental data (back-calculated) used is from Ref.\ \cite{Stolwijk04} but superimposed with a weak noise of $\pm 5$ \%. The standard deviation is shown by the vertical bars. Parameters used for the different sets are listed under Tab.\ \ref{tabset}.}
\end{figure}

Fig.\ \ref{DiffSO}(a) shows the values of diffusivities of the free Na$^+$ (dotted line) and
free I$^-$ (solid line) obtained from the solution of Eq.\ \eqref{SOmod} with
the parameters SET I.
It can be observed that $D_{I,f}$ is larger than
$D_{Na,f}$ over the temperature range shown.
The curves monotonically decrease from the high to the low temperature end. 
$D_{I,f}$ exceeds $D_{Na,f}$  by about a factor of about 6$-$7 at the low temperature end
to about 2$-$3.5 at the high temperature end. 
Particularly, at T = 450 K, $D_{I,f}/D_{Na,f} = 3.6$ which is about a factor of 1.8 larger 
than what is seen in the simulation (i.e.\ $D_{I,f}/D_{Li,f} \approx 2.0$). 
However, this factor is expected to depend upon the chain length of the polymer host
as shown in Ref.\ \cite{Hayamizu1} in agreement with the current observations. By only supplanting 
the parameters $\eta_{Na}=\eta_I=1.0$
in SET I
and letting the others unaltered (i.e.\ SET II) one observes a strong 
shift by a factor of $\approx$ 3 in $D_{I,f}$ (dashed line) to larger values
especially at the high temperatures and by about 20 \% at the low temperatures.
By choosing the parameters 
from SET III, $D_{I,f}$ decreases less than 20 \% at high temperatures and even less at low temperatures   
[see Fig.\ \ref{DiffSO}(b)] when compared against SET II. 
Choosing the values $\alpha=74$ and $\beta=\gamma=\chi=\eta_{Na}=\eta_I=1.0$ (i.e.\ SET IV)
would have reproduced the results $D_{I,f}/D_{Na,f} \approx 8$ obtained by Stolwijk-Obeidi (not shown)
at all temperatures.

The dynamics of the free Na$^+$ ions
exhibits insignificant alteration between the parameter sets I, II and III. This can be explained
from Eq.\ \eqref{SOmod}(b) under the condition $\alpha \approx 1$ which gives $D_{Na,f} \approx D_{Na}^{tr}$
and therefore making $D_{Na,f}$ relatively independent of the parameters. 

In the foregoing, we have identified $\eta_{Li}$ and $\eta_{I}$ as important players 
in the estimation of $D_{I,f}$ while applying the extended SO
model to our simulation data.
Provided that \mbox{$\alpha \approx 1$}, as is the case for the polymer electrolyte under investigation, 
a change in the other parameters ($\beta$, $\gamma$ and $\chi$)
are not as important in generating the large change in the diffusivity of the free I$^-$ ions
at the high temperatures.   
We, next, show that the reason for the large sensitivity of $D_{I,f}$ originates 
primarily from the small changes in $\eta_I$ whereas the influence of $\eta_{Na}$
in this case is rather weak. If $\alpha \approx 1$ then $r_{I,nf} \propto 1/\eta_{Na}$ (from Eq.\ \eqref{rLinf} and Eq.\ \eqref{SOmod}(a)) and $D_{I,nf} \propto \eta_{Na}/\eta_{I}$ (see Appendix II) which 
implies that $r_{I,nf} D_{I,nf} \propto 1/\eta_{I}$. From Eq.\ \eqref{SO}, $ D_{I,f} = \frac{D_{I}^{tr} - r_{I, nf} D_{I, nf}}{1-r_{I,nf}} $, and if $r_{I,nf}$ is large then the second term in the numerator of the rhs dominates and the denominator is also correspondingly small. Therefore, 
a small decrease in $\eta_I$ can cause a large increase in $D_{I,f}$ as observed in Fig.\ \ref{DiffSO}(a) at high temperatures where the fraction of non-free ions is large (see below). In contrast, at lower temperatures the proportion of non-free I$^-$ ions drops and thereby stabilizing
 $D_{I,f}$ from small variations in $\eta_I$.

Fig.\ \ref{FrNonFreeSO}(a) shows the corresponding
fraction of non-free ions for the sets I, II and III. Similar to the plots
of diffusivities of the free ions, one finds monotonously decreasing and lightly bent behavior
over the entire temperature range. 
Moving from SET I to SET II one observes an increase of $r_{nf}$ by about 6 \% which can be 
explained from the proportionality: $r_{nf} \propto 1/\eta_{Na}$.
In contrast, while changing parameters from SET I to SET III 
there is only a inconspicuous increase
in the proportion of non-free ions by less than 2 \%. This is due to the nullifying effect of 
the simultaneous variation of the parameters between the two sets of parameters and thus, appearing as 
similar fractions of non-free ions which is rather a mere coincidence.
Choosing SET IV, however, results in a scanty fraction of non-free ions (see Fig.\ \ref{FrNonFreeSO}(b)).

\begin{figure}
\centering
  \includegraphics[clip,width=0.5\linewidth]{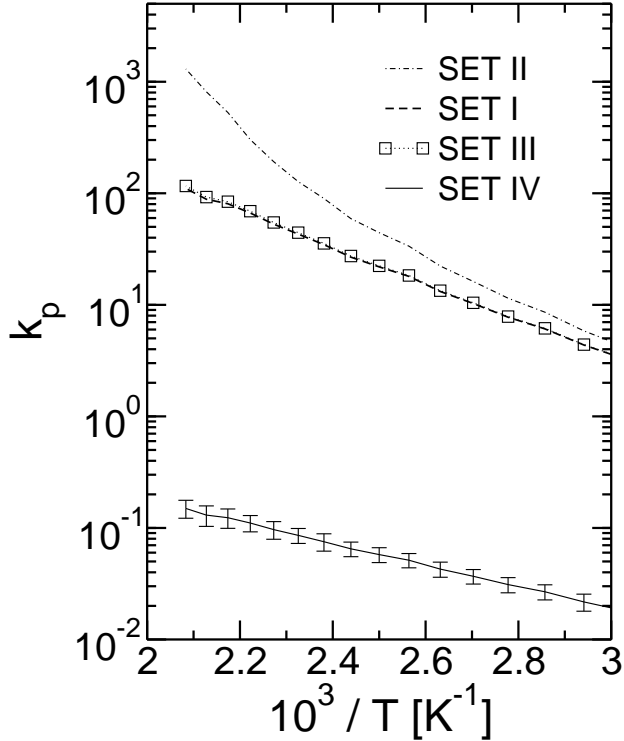}
  \caption{\label{kpAllmodels} Pair formation constant $k_p$ with parameters from the different sets 
listed under Tab.\ \ref{tabset}.}
\end{figure}
Finally, the ion-pairing reaction constant, $k_p = r_{nf}/(1 - r_{nf})^2$ [see Ref.\ \cite{Stolwijk04,Stolwijk07}] for the equilibrium reaction Na$^+$ + I$^-$ $\rightleftharpoons$ Na$^+$I$^-$ at the different temperatures is analyzed. Note that the above equation of $k_p$ is strictly applicable only if the non-free ions exist as ion-pairs. These are shown in Fig.\ \ref{kpAllmodels} for the four cases mentioned. 
For the original SO model (SET IV) the $k_p$ values are orders of magnitude less than those obtained 
from the other sets primarily due to the $\alpha$ parameter.
A fit with $k_p = k_{p0} \exp(-\Delta H_p/k_BT)$ where $k_{p0}$ is the prefactor,
$\Delta H_p$ is the formation energy of the pair and $k_B$ is the Boltzmann constant yields the energetic term. This is listed in Tab.\ \ref{tabfit} along with the fraction of non-free ions at T = 450 K compiled for the different cases and sources. 
$\Delta H_p$ calculated from the simulation using $r_{nf}$ at T = 450 K is consistent with that obtained from the experiment by using the extended SO model with the parameters in SET I. Comparing the fraction of non-free ions between the original SO model \cite{Stolwijk04} ($r_{nf}=0.08$) against the extended SO model with SET I ($r_{nf} = 0.88$) the discrepancy is evident at T = 450 K. In fact for the former a high ion-pair diffusivity is required to match the small fraction of ion-pairs in order to generate a reduction of the average ionic diffusivity by a factor of 3 to be compatible with the observed collective diffusivity. Note that with SET II (i.e.\ by neglecting the importance of $\eta_L$) one would have been tempted to erroneously conclude that the pair formation is not governed by a single $\Delta H_p$. As an additional information, the pair formation energy determined from the Raman spectroscopic work on poly(propylene oxide)PPO/NaCF$_3$SO$_3$ with O:Na = 30:1 by Kakihana et al.\ \cite{Kakihana90} provides $\Delta H_p = 0.16$ eV but with a high pair fraction. Of course, it would have made an interesting comparison if spectroscopic data had been available for the PEO/NaI system. \\

\begin{table}
  \caption{\label{tabfit} Comparison of the energetic values obtained from fits of $k_p(T)$ from different sources. The salt concentration in terms of the ratio of ether oxygen atoms to cations is included in the parenthesis.}
  \begin{tabular}{|l| c| c| c| }\hline  
  System & Remark  &  $\Delta H_p$ [eV] & $  r_{nf}$ \\
  & & &    (T = 450 K)    \\  \hline 
  PEO/LiI (20:1) &  Simulation                &   0.31     &  0.92 \\ \hline
  PEO/NaI (30:1) &  Params, SET I    &  0.33  &  0.88  \\
          & (Tab.\ \ref{tabparams}) & & \\ \hline
  PEO/NaI (30:1) &  Ref.\ \cite{Stolwijk07}      &  0.19 & 0.08  \\ \hline
  PPO/NaCF$_3$SO$_3$ (30:1) & Ref.\ \cite{Kakihana90}      &  0.16 & $>$ 0.70  \\ \hline
  \end{tabular}
\end{table}

\section{\bf Summary}

In the first section ionic correlations in a polymer electrolyte (e.g.\ the archetypal PEO/LiI) stemming from ionic associations are elucidated by constructing elementary correlation functions. One of these functions included the time-evolution of the directional correlation of a pair of ions (both like-charged and unlike-charged) in dependence of the separation distance between the ions. The correlations from the \hyp{Li}{I} pairs are found to be the strongest in comparison to the \hyp{Li}{Li} and \hyp{I}{I} pairs of ions. Furthermore, the correlation effects are found to exist beyond the nearest neighborhood of counterions, albeit reduced due to the screening of interactions. In essence, the strong structural correlations together with the non-trivial values of the directional correlations between the cation-anion pairs lead to a reduction of the ionic conductivity of the polymer electrolyte compared to the average diffusivity of the ions. The joint mobility of a pair of ions calculated as the product of the magnitude of the displacement vectors of the ions showed that during a certain time this is almost independent of the structural correlations between the ions. One can thus infer that the dynamics of the different states of ions (i.e.\ free ions, ion-pairs, ion-triplets etc.) will be similar. Intuitively, due to the fact that most of the lithium ions are coupled to the ether oxygen atoms of the PEO one would expect some correlations between the dynamics of the free and the associated ions. This is, additionally, verified by computing the mean square displacement of the free Li$^+$, isolated Li$^+-$I$^-$ pairs and the isolated I$^--$Li$^+-$I$^-$ triplets. This is in stark contrast to the conclusion from the Stolwijk-Obeidi model \cite{Stolwijk04} which predicted the dynamics of ion-pairs to be orders of magnitude faster than the free ions.

In the second section the SO model is extended by starting from the Einstein-like equation which describes the collective diffusivity of the ions from the single-particle dynamics of the ions (the self terms) and the correlations from the distinct pairs of ions. Instead of describing the overall correlation effects only in terms of the dynamics and the fraction of isolated ion-pairs we generalised the approach by considering the dynamics and the fraction of non-free ions. Typically, the non-free ions account for the presence of ion-triplets or any higher order clusters along with the ion-pairs. Additionally, various approximation parameters, $\chi_i$, are incorporated that keep track of the simplifications made to the correlated term of the Einstein-like equation. The simplifications can further be understood in terms of simple microscopic pictures. The parameters could be extracted from simulations and are found to be close to unity which is a consequence of the nature of ionic association and dynamics in polymer electrolytes. If each of these parameters except $\alpha$ (which appeared as a result in the SO model) are chosen to be exactly unity one recovers the SO model. The extended SO model along with the inference of dynamical similarity between the free and the non-free ions is applied to the same data as in Ref.\ \cite{Stolwijk04} but with the additional parameters chosen from simulation.
It is found that the fraction of non-free ions are high and consistent with the simulation results. The parameter $\eta_{I}$ also turns out to be more important compared to the other parameters ($\eta_{Na}$, $\beta$, $\gamma$, $\chi$) specifically for yielding the estimates of the diffusivity of free I$^-$ ions. Typically, $\eta_{Na}/\eta_{I} \ne 1$ expresses the difference in timescales over which the cations and the anions become diffusive and reflects a characteristic of polymer electrolytes where the cations are predominantly complexed with atoms/groups located along the main backbone of the polymer for larger timescales in comparison to the anions which are coordinated to the same cationic neighbors for relatively short times. It must be, however, borne in mind that the results (i.e.\ fraction of non-free ions and the diffusivity of free ions) from the SO model depend on the accuracy of the input experimental data and possibly on the temperature dependence of one or more of the parameters (e.g.\ $\eta_{Na}$, $\eta_I$).
Further work is required to map the values of the parameters as a function of interaction between ions and the resulting collective diffusivity of ions that would impart general applicability to quantifying ionic association.
\newline\newline
{\bf Appendix I}\\
Let $n$ be the total number of ions in the system and so the number of cations and anions are $n/2$ respectively. Defining the quantity $Z$ as follows
\begin{equation}
Z = \sum_i^{n/2} \langle \delta_{s(t_0),f} +
\delta_{s(t_0+t), f} + \delta_{s(t_0), nf} + \delta_{s(t_0+t), nf} \rangle_{t_0} = n
\end{equation}
one can show that Eq.\ \eqref{MSDion} can be derived from Eq.\ \eqref{specMSD} and Eq.\ \eqref{specr1} through the following steps:

\begin{subequations}
\begin{align}
\langle \Delta R^2_{Li}(t) \rangle &= \frac{\sum_i \langle \{ \delta_{s(t_0), f} + \delta_{s(t_0+t), f} + \delta_{s(t_0), nf} + \delta_{s(t_0+t), nf} \} \Delta R^2_{i}(t) \rangle_{t_0} }{Z} \\
 &= \frac{\sum_i \langle \{ \delta_{s(t_0), f} + \delta_{s(t_0+t), f} \} \Delta R^2_{i}(t) \rangle_{t_0} }{Z}
 +
\frac{\sum_i \langle \{ \delta_{s(t_0), nf} + \delta_{s(t_0+t), nf} \} \Delta R^2_{i}(t) \rangle_{t_0} }{Z} \\
 &=    \frac{\sum
_i \langle \delta_{s(t_0), f} + \delta_{s(t_0+t), f} \rangle_{t_0}}{Z} \bullet \frac{\sum_i \langle \{ \delta_{s(t_0), f} + \delta_{s(t_0+t), f} \} \Delta R^2_{i}(t) \rangle_{t_0} }{\sum_i \langle \delta_{s(t_0), f} + \delta_{s(t_0+t), f} \rangle_{t_0}} \\
 &+ \frac{\sum                       _i \langle \delta_{s(t_0), nf} + \delta_{s(t_0+t), nf} \rangle_{t_0}}{Z}  \bullet \frac{\sum_i \langle \{ \delta_{s(t_0), nf} + \delta_{s(t_0+t), nf} \} \Delta R^2_{i}(t) \rangle_{t_0} }{\sum
_i \langle \delta_{s(t_0), nf} + \delta_{s(t_0+t), nf} \rangle_{t_0}} \\
&= r_{Li, f} \: \langle \Delta R^2_{Li,f}(t) \rangle  +  r_{Li, nf} \: \langle \Delta R^2_{Li,nf}(t) \rangle .
\end{align}
\end{subequations} 
\newline
{\bf Appendix II}\\
Here, we show how to rewrite Eq.\ \eqref{SO} (for $L = Li$ and $L =I$) and Eq.\ \eqref{SO3}.
The fraction of non-free Li$^+$ ions can be determined as : $ r_{nf} = 0.5(r_{Li, nf} + r_{I,nf}) = 0.5 (r_{Li,nf} + \gamma r_{Li, nf}) = 0.5 r_{Li, nf}(1 + \gamma)$. Therefore,
\begin{equation}\label{rLinf}
r_{Li, nf} = 2 r_{nf} \frac{1}{1+\gamma} \: \mbox{and} \: r_{I, nf} = 2 r_{nf} \frac{\gamma}{1+\gamma}.
\end{equation}
The corresponding proportion of free Li$^+$ and I$^-$ ions are given by $r_{Li,f} = 1 - r_{Li, nf}$ and $r_{I,f} = 1- r_{I, nf}$. The solutions given in Eq.\ \eqref{SOmod} can
be obtained from the following equations:
\begin{subequations}
\begin{align}
\begin{split}\label{Ap_b}
D_{Li}^{tr} &= r_{Li, f} D_{Li, f} + r_{Li, nf} \alpha D_{Li, f} = D_{Li,f} (r_{Li,f} + \alpha r_{Li, nf}) \\
            &= D_{Li,f}\left[1 - (1-\alpha) \frac{2r_{nf}}{1+\gamma} \right]
\end{split}\\
\begin{split}
D_I^{tr} &= r_{I,f} D_{I,f} + \alpha \beta \frac{\eta_{Li}}{\eta_I} r_{I,nf} D_{Li, f} \\
         &= D_{I,f}[1 - \gamma \frac{2r_{nf}}{1 + \gamma}] +  \alpha \beta \frac{\eta_{Li}}{\eta_I} \frac{2 r_{nf}}{1 + \gamma} \gamma D_{Li,f}
\end{split}\\
\begin{split}
2 D_{cross} &= 2 \chi \frac{r_{nf}}{1 + \gamma} (\eta_{Li} D_{Li,nf} +  \gamma \eta_I D_{I,nf}) = 2 \chi \frac{r_{nf}}{1+\gamma}(\eta_{Li} \alpha D_{Li,f} + \gamma \eta_I \alpha \beta \frac{\eta_{Li}}{\eta_I} D_{Li,f}) \\ &= 2\chi \frac{r_{nf}}{1+\gamma} \eta_{Li} (1 + \gamma \beta) \alpha D_{Li,f}
\end{split}
\end{align}
\end{subequations}
where we have used Eq.\ \eqref{gambetalp} and specifically
in Eq.\ \eqref{Ap_b} we have used $D_{I,nf} = \beta \frac{\eta_{Li}}{\eta_I} D_{Li,nf} = \alpha \beta \frac{\eta_{Li}}{\eta_I} D_{Li,f}$. Straightforward manipulation then gives rise to Eq.\ \eqref{SOmod}(a-c).
\newline\newline
{\bf Acknowledgement} \\
We thank M.\ Sch\"onhoff, N.\ A.\ Stolwijk and M.\ Vogel for several helpful comments and discussion. This work is carried out within the framework of {\it Sonderforschungsbereich 458} and supported by the NRW Graduate School of Chemistry, M\"unster.


\begin{thebibliography}{31}
\bibitem[{not()}]{note4}
\bibinfo{note}{The collective diffusivity $D_\sigma^\prime$ in the Ref.\
  \cite{Stolwijk04} have an additional factor of 2, i.e.\ $D_\sigma^\prime = 2
  D_\sigma$ where $D_\sigma$ is the collective diffusivity defined in this
  article.}
\bibitem[{int()}]{interact}
\bibinfo{note}{The interactions involving the I$^{-}$ in Tab.\ 4 and 6 of ref.\
  \cite{PEOLiIFF97} were modified as follows: $A(I-I) = 28881$ kcal/mol,
  $B(I-I)=1.75$ \AA$^{-1}$, $C(I-I)=0$; $D(I-I)=.71 \times 3453$ \AA$^{6}$
  kcal/mol , $C(O-I)=0$. All $D$ parameters were scaled by a factor of 0.71.
  Additionally, the equation of the force involving the $D$ term was scaled by
  a distance dependent scaling factor $\epsilon_b \exp (-\frac{a}{r}
  \ln\epsilon_b)$ beyond 3.6 \AA\ ($\equiv a$) with $\epsilon_b=100$. The force
  fields were then reasonably fitted with Buckingham potential in the present
  work.}
\end{thebibliography}

\end{document}